\documentclass[aps,prd,showpacs,nofootinbib,11pt]{revtex4-1}
\usepackage[utf8x]{inputenc}
\usepackage{amsmath,amsfonts,amssymb,graphicx,accents,mathrsfs,mathtools,cleveref,boxedminipage}
\usepackage{braket,textcomp}
\usepackage[percent]{overpic}
\usepackage{cleveref,braket,textcomp}
\usepackage[percent]{overpic}
\usepackage{color}
\usepackage[makeroom]{cancel}
\usepackage{cleveref}
\def\beq{\begin{eqnarray}}
\def\eeq{\end{eqnarray}}

\def\nn{\nonumber\\}

\DeclareMathOperator{\sech}{sech}

\begin{document}
\title{\boldmath Inhomogeneous Jacobi equation and Holographic subregion complexity}
\author{Avirup Ghosh}\email{avirup.ghosh@iitgn.ac.in}
\affiliation{Indian Institute of Technology, Gandhinagar, 382355, Gujarat , India.}
\author{Rohit Mishra,}\email{mrohit996@gmail.com}
\affiliation{School of Physical Sciences, Indian Association for the Cultivation of Science,
2A $\&$ 2B Raja S.C. Mullick Road, Kolkata 700032, India}

\begin{abstract}
We derive a general expression for obtaining Holographic subregion complexity for asymptotically $AdS$ spacetimes, pertubatively around pure $AdS$ using a variational technique. An essential step in finding subregion complexity is to identify the bulk minimal surface of the entangling subregion. Our method therefore heavily relies on solutions of an inhomogeneous version of Jacobi equation, used to study deformations of the entangling surface for perturbations of the bulk metric. Using this method we have obtained the change in complexity for a strip and a circular disk like subsystem for \emph{boosted} black brane like perturbations over pure $AdS_4$. As a corollary, we find that for spherical subsytems in $3+1$ dimensional bulk, the linear change of subregion complexity for \emph{boosted} black brane like perturbations over pure $AdS_4$ , vanishes.
\end{abstract}
\maketitle
\section{Introduction}
In classical computation the notion of complexity is a characterization of how hard a computational problem is? A sorting problem for example roughly requires $N(N-1)$ steps to complete, where $N$ is the no. of elements to be sorted. In the the asymptotic limit this is $\mathcal O(N^2)$ and the problem is therefore said to be of polynomial complexity. On the other hand a prime divisor problem is believed to require steps, which grow exponentially with the no. of digits. The problem is then termed to be of exponential complexity. Another, similar characterization of complexity is circuit complexity, which is a measure of the number of elementary gates required to go from one `classical' input to a `classical' output. An important corollary of such a finding is the fact that more complex a problem, the better a computer modeled along that problem, will be. 

In the quest for a quantum computer one therefore needs a notion of complexity for quantum computations. In a quantum computer a classical bit is replaced by a qubit which is a two dimensional Hilbert space. A computation is then the action of a unitary operator on the states of this Hilbert space. A gate in quantum computation is just a unitary operator acting on a state. The notion of circuit complexity can then be suitably modified as the number of elementary unitary operations required to go from a given input state to another output state. An operational definition for this was given in \cite{nielsen2005geometric}, following which the circuit complexity in quantum field theories was calculated in \cite{Jefferson:2017sdb}.

Considering the fact that calculating complexity in QFTs might be a difficult job, one considers the question of whether a simple gravitational dual of such a quantity exists for states which have a gravitational dual in the Holographic context. The proposal for such a quantity was first given in an attempt to establish the ER=EPR paradigm in \cite{Susskind:2014rva}. A global AdS Schwarzschild black hole is described by two copies of CFT living on the two asymptotic ends of the geometry. While each copy of the CFT is in a thermal state, the state can equally be described by a thermofield double, which is a pure but entangled state in the product of these two Hilbert spaces. One can then assume that a measurement is done in one of the copies of the two Hilbert spaces. The complexity of this measurement affecting the state in the other Hilbert space was conjectured to be the volume of an extremal slice connecting two constant time slices, each in one of the asymptotic boundaries. This later came to be known as the $Complexity=volume$ conjecture. To avoid certain shortcomings in the above conjecture the $Complexity=action$ was put forward in \cite{Brown:2015bva}. The conjecture states that complexity is equal to the action evaluated in the domain of dependence (in the bulk) of the two asymptotic constant time slices. An essential improvement of this over the previous one was that it fixes an arbitrary multiplicative constant in the $Complexity=volume$ conjecture. Recently by identifying the symplectic two form on the phase space of bulk gravity gravity with that of the symplectic structure on the projective space of states of the boundary field theory \cite{Belin:2018fxe} it has been shown that a slight modification of Nielsen's prescription is indeed dual to the $Complexity=volume$ prescription \cite{Belin:2018bpg}.

But this prescription holds for the state of the boundary subregion as a whole. A quantum computation may proceed by acting unitary operators on subregions of the full system. The state of the full system is then built by such operations performed one at a time. Hence it is sensible to define an analogous quantity for subregions. A working definition for this was given in \cite{Alishahiha:2015rta}. The idea can be elaborated as follows. Suppose one considers a subregion $A$ (defined by a spatial subregion of the boundary) of the boundary. One then needs to find a minimal co-dimension two surface $\mathcal S$ in the bulk that is homologous to $\partial A$. Recall that the area of this surface is nothing but the entanglement entropy of $A$. The subregion complexity is then the volume of a maximal co-dimension one surface bounded by $\mathcal S$ and $A$.

In this paper we will be mainly interested in this $Complexity=volume$ definition for subregion. It is important to note that there are very few exact solutions for the minimal surface equation of codimension one/two surfaces in the asymptotically $AdS$ spacetimes. For static spacetimes foliated by $t$ say, it turns out that both $\mathcal S$ and $\Sigma$ lie on a $t=constant$ co-dimension one slice and the problem is, fairly well, tractable. But once one introduces rotation, the calculation becomes difficult. This is also the case for time dependent geometries. The reason for the complication is that $\mathcal S$ and $\Sigma$ no more lie on the $t=constant$ (here $t$ is the foliating time for the boundary field theory) slices. One then resorts to perturbative schemes in which complexity/entanglement entropy is calculated by studying perturbations over $AdS$, for which the relevant surfaces are known. There are two approaches to do this. The first approach is a perturbative one, which requires solving the minimal surface equation perturbatively in some small approximation \cite{bender2013advanced,lin1988mathematics}. One then uses this approximate solution to evaluate the area or volume order by order\cite{Bhattacharya:2012mi,Mishra:2015cpa,Quijada:2017zif, Bhattacharya:2019zkb}. The second approach is a variational one. It starts by taking the variation of the area and volume functional that incorporates both changes in embeddings and metric perturbation \cite{Lewkowycz:2018sgn,Mosk:2017vsz,Bao:2019bib,Engelhardt:2019hmr,Speranza:2019hkr}. The changes in the embedding is obtained by solving an inhomogeneous Jacobi equation, the inhomogeneity of which is sourced by metric perturbations. It can be shown that in some known cases, the first and second variation of the area functional, obtained using the variational approach, exactly matches with the first and second order results, obtained  using the perturbative approach \cite{Ghosh:2017ygi,Ghosh:2016fop}. In this paper we extend this variational approach from area functional to the volume. It turns out that the differential equations that one  
needs to solve, for calculating subregion complexity are all of second degree and second order. The first for finding out the surface $\mathcal S$ that determines the entanglement entropy and the other for finding out the maximal slice $\Sigma$ (say). 
 
In this paper we provide a general systematic scheme for performing the variational expansion. Since the definition of subregion complexity requires finding the bulk entangling surface of the subregion, the scheme heavily relies on finding how the entangling surface deforms under the perturbation of the bulk geometry. A definite procedure for finding such deformations was given in \cite{Ghosh:2016fop,Mosk:2017vsz, Ghosh:2017ygi}. They are given by a vector field which has support on the entangling surface $\mathcal S$ and arise as a solution of a linear elliptic differential equation, which is nothing but an inhomogeneous version of the well known Jacobi equation for minimal surfaces. We find that the first order change in the holographic subregion complexity is nothing but the integral of a component of this vector field over $\mathcal S$. Using the solutions obtained in \cite{Ghosh:2016fop,Ghosh:2017ygi} we compute the first order changes of complexity for $BTZ$ like perturbations over $AdS_{2+1}$ and boosted black brane like perturbations over $AdS_{3+1}$.  

We also derive general expressions for finding the second variation of complexity. As, such a calculation requires a separate discussion, we do not explicitly solve any specific examples for this case. However we do point out the advantages of finding such general expressions. In particular, how these can be put to use in the context of both holographic complexity as well as entanglement entropy, is discussed. In fact the general expressions obtained would allow one to calculate the change in complexity as well as entanglement entropy for variations of the subregion as well, i.e in cases where the boundary state does not change, rather the boundary subregion is deformed. 
 
\section{Notations and conventions}\label{N&C}
The theory of perturbation to be used here will essentially be the one in the covariant framework introduced in \cite{Stewart:1974uz}. The perturbation of $d+1$ dimensional space time $(\mathcal M,{g})$ to another $d+1$ dimensional space time $(\mathcal M',g')$ is characterized by a differentiable map $\Phi:\mathcal M\rightarrow \mathcal M'$ which is however not isometric.  $(\mathcal M',g')$ is then called a perturbation over $(\mathcal M,g)$ if $\accentset{(1)}{P}=\Phi_*g'-g$ is a small perturbation over $g$.

Consider a surface $\mathscr S$ isometrically embedded in $\mathcal M$ and given by the function $f:S\rightarrow \mathcal M$. Let the coordinates on $\mathscr S$ be denoted by $\tau^a$. If $\mathscr S$ has a boundary $\partial \mathscr S$ then let its coordinates be denoted by $\chi ^A$. This will be the general strategy followed in the text. While discussing an embedded surface, its coordinates will be denoted by $\tau^a$ and the coordinates on its boundary will be denoted by $\chi^a$. The ambient spacetimes coordinates will always be denoted by $x^\mu$. The induced metric on $\mathscr S$ is given by $h=f_*~g$, which, in the local coordinates can be written as $h_{ab}=g(\partial_a,\partial_b)=\frac{\partial x^\mu}{\partial\tau^a}\frac{\partial x^\nu}{\partial\tau^b}g(\partial_\mu,\partial_\nu)$. The quantity $\frac{\partial x^\mu}{\partial\tau^a}\partial_\mu$ is the push forward of the purely tangential vector field $\partial_a$ to $\mathcal M$. `$h_{ab}$' is the first fundamental form on $\mathscr S$.  One can decompose the tangent space at the point $x~\in~\mathscr S$ into the tangent space of $\mathscr S$ and the space of normal vectors as $T_x\mathcal M=T_x\mathscr S\oplus T_x^{\perp}\mathscr S$.  Let the covariant derivative on $\mathscr S$ be denoted by $D: T\mathscr S \otimes T\mathscr S\rightarrow T\mathscr S$. Let $X,Y~\in T\mathscr S$. Then the Gauss decomposition allows us to write,
\begin{gather}
\nabla_XY=D_XY+K(X,Y),
\end{gather}
where $D_XY$ is purely tangential and $K(X,Y)$ is a vector in the normal bundle and is the extrinsic curvature or the second fundamental form. A connection $\nabla^{\perp}_XN^\perp$ in the normal bundle can be defined as $\nabla^\perp:T\mathscr S\otimes T^\perp \mathscr S\rightarrow T^\perp \mathscr S$, where $X\in T\mathscr S$ and $N^\perp\in T^\perp\mathscr S $. The shape operator $W_{N^\perp}(X)$ is then defined as,
\begin{gather}
\nabla_XN^\perp=\nabla_X^\perp N^\perp-W_{N^\perp}(X).
\end{gather} 
The Weingarten equation then relates the shape operator and the extrinsic curvature,
\begin{gather}
g(W_{N^\perp}(X),Y)=g(N^\perp,K(X,Y)),
\end{gather}
where $X,Y\in T\mathscr S$ and $N^\perp\in T^\perp\mathscr S$. The Riemann tensor is defined as,
\begin{gather}
R(W,U)V:=[\nabla_W,\nabla_U]V-\nabla_{[W,U]}V
\end{gather}
Similarly one can define the Riemann tensor on $\mathscr S$ as,
\begin{gather}
\mathcal R(X,Y)Z:=[D_X,D_Y]Z-D_{[X,Y]}Z
\end{gather}
The Gauss and Codazzi equations relates the intrinsic Riemann tensor and the spacetime Riemann tensor. Let $X,Y,Z,W\in T\mathscr S$ and $N^\perp\in T^\perp\mathscr S$. Then the Gauss equation is given as,
\begin{gather}
g(R(X,Y)Z,W)=g(\mathcal R(X,Y)Z,W)-g(K(X,Z),K(Y,W))+g(K(X,W),K(Y,Z)),
\end{gather}
and the Codazzi equation as,
\begin{gather}
g(R(X,Y)N^\perp,Z)=g((\nabla_YK)(X,Z),N^\perp)-g((\nabla_XK)(Y,Z),N^\perp)
\end{gather}
Due to the presence of perturbations a variation will have two contributions, one which is a flow along a vector $N\in T\mathcal M$, and another variation $\delta_g$ which is purely due to metric perturbations. The metric perturbation will be given by,
\begin{gather}
(\delta_g~g)(\partial_\mu,\partial_\nu):=\bigg[\Phi_*~g'-{g}\bigg](\partial_\mu,\partial_\nu)=\accentset{(1)}{P}(\partial_\mu,\partial_\nu),~~~(\delta_g^2g)(\partial_\mu,\partial_\nu)=(\delta_g \accentset{(1)}P)(\partial_\mu,\partial_\nu):=\accentset{(2)}{P}(\partial_\mu,\partial_\nu)
\end{gather}
where $\accentset{(1)}{P}$ is a symmetric bilinear form on $\mathcal M$. Note that $\delta_g$ only acts on the metric and does not change the vector fields $\partial_\mu$. Now suppose there is a covariant derivative $\nabla'$ in $\mathcal M'$ compatible with $g'$, then for $X,Y\in T\mathcal M$,
\begin{gather}
\accentset{(1)}{C}(X,Y):=\delta_g\bigg(\nabla_XY\bigg)=\tilde\nabla_XY-\nabla_XY,
\end{gather}
where $\tilde\nabla=\phi^*\nabla'$ is the pullback connection on $M$. It is important to note that $C(X,Y)$ is a vector field on $\mathcal M$.
\begin{gather}
\delta_g \accentset{(1)}{C}(X,Y)=\accentset{(2)}{C}(X,Y)-2 \accentset{(1)}{P}(\accentset{(1)}{C}(X,Y)),
\end{gather}
where $\accentset{(1)}{P}(X)$ is a vector defined as $g(\accentset{(1)}{P}(X),Y)=\accentset{(1)}{P}(X,Y)$.
 \section{Variations on general minimal surfaces}\label{VGM}
 \subsection{Variation of area}
 In this section we will calculate the variation of area. These expressions are in general true for any minimal submanifold. Note that these have been obtained before, but certain boundary terms have been ignored \cite{Mosk:2017vsz,Ghosh:2017ygi}. We will retain the boundary terms, obtained in the process, and later see how these boundary terms are useful. Only the first order expressions will be used to find the perturbative change of holographic subregion complexity for boosted black brane like perturbations over pure $AdS$  in section (\ref{FOV}). The second order expressions will however not be evaluated for the examples taken but the usefulness of these expressions in obtaining properties of subregion complexity and holographic entanglement entropy will be discussed. We will also discuss the scheme which one must follow in order to evaluate the second order variation, given a perturbation. 
 
Let us consider a one parameter flow (generated by the vector field $N$) of the area and an $n$-dimensional minimal surface $\mathscr S$ embedded in a $d+1$ dimensional spacetime. First variation of area of the submanifold is given by,
 \begin{gather}
 \delta_{N} A=-\int_{\mathcal S}g(N^\perp,K(\partial_a,\partial_b))h^{ab})~\sqrt{h}d^n\tau+\frac{1}{2}\int_{\mathscr S} \accentset{(1)}{P}(\partial_a,\partial_b)h^{ab}\sqrt{h}~d^{n}\tau-\int_{\partial\mathscr S} g(\hat r, N^T)\sqrt{\gamma}~d^{n-1}\chi,\label{c}
 \end{gather}
 where `$\tau^a$' are coordinates on $\mathscr S$ and `$\chi^A$' on ${\partial\mathscr S}$. $h_{ab}$ and $\gamma_{AB}$ are the induced metrics on $\mathscr S$ and $\partial\mathscr S$. The unit co-normal to ${\partial\mathscr S}$ w.r.t $\mathscr S$ is denoted by $\hat r$.
 $K(\partial_a,\partial_b)$ is the extrinsic curvature of $\mathscr S$ in $\mathcal M$. $\perp$ and $T$ denotes component perpendicular and parallel to $\mathscr S$, respectively. Further $N$ has been taken to commute with $\partial_a$'s. The first term in eq. (\ref{c}) is however zero for a minimal surface. While deriving the second variation we will differentiate the above expression rather than the one obtained after imposing the minimal surface condition. Thus we have,
 
 \begin{gather}\label{sc}
 \delta^{2}_NA=\int_{\mathscr S} d^{n}\tau~\sqrt{h} \Bigl(h^{ab}h^{cd}P(\partial_b,\partial_d)g(N^{\perp},K(\partial_a,\partial_c))-h^{ab}g(C(\partial_a,\partial_b),N^{\perp})\Bigr)\\\notag
 +\int_{\mathscr S} d^{n}\tau~\sqrt{h}\Bigl[{h^{ab}\over 2}\accentset{(2)}{P}(\partial_{a},\partial_{b})-{1\over 2}h^{ac}h^{bd}\accentset{(1)}{P}(\partial_{a},\partial_{b})\accentset{(1)}{P}(\partial_{c},\partial_{d})+{1\over 4}h^{ab}h^{cd}\accentset{(1)}{P}(\partial_{c},\partial_{d})\accentset{(1)}{P}(\partial_{a},\partial_{b})\Bigr]\\\notag
 -\int_{\partial\mathscr S} d^{n-1}\chi~\sqrt{\gamma}\hat{r}_a[h^{ab}h^{cd}\accentset{(1)}{P}(\partial_c,\partial_d)g(N,\partial_b)]-2\int_{\partial\mathscr S} d^{n-1}\chi~\sqrt{\gamma}\hat{r}_a[h^{ab}\accentset{(1)}{P}(N^{\perp},\partial_b)]\\\notag
 -\int_{\partial\mathscr S} d^{n-1}\chi~\sqrt{\gamma}\hat{r}_a[h^{ab}g(N^{\perp},{\nabla^{\perp}}_{\partial_{b}}N^{\perp})]
 -\int_{\partial\mathscr S} d^{n-1}\chi~\sqrt{\gamma}g((\nabla_{N}N)^T,\hat{r}),
 \end{gather}
 The details of this calculation can be found in appendix \ref{sva}. In section \ref{pscheme} we will discuss how to evalute individual terms in the above expression. In other words we will discuss what decides the individual terms arising in the above expression. Before delving into such a discussion we will first derive the second order perturbation equation. This will help us to discuss the origin of terms like $(\nabla_NN)^T$ in the above expression.

 \subsection{Second order perturbation equations}\label{sop}
 \subsubsection{First order perturbation revisited}
 Assume that under a perturbation of the ambient metric the minimal surface $\mathscr S$ deforms to a minimal surface $\mathscr S'$. Let the deviation be denoted by the vector field $N$.
 We will recall the derivation of the first order inomogeneous Jacobi equation, for $\mathscr S$, to facilitate further discussion. Finding the first order Jacobi equation amounts to equating the total variation of the mean curvature vector $H$ to zero, i.e,
 \begin{equation}
 \mathcal L_N H+\delta_g H=0
 \end{equation}
 However note that $\mathcal L_N H =\nabla_N H-\nabla_H N $. Since $H=0$ on the back-ground one only needs to calculate $\nabla_N H$. We will see that this will also be the case for the second order perturbation equations. Let us start by calculating $\nabla_N\nabla_{\partial_a}\partial_b$. $N$ will be taken to commute with all the $\partial_a$'s. However note that neither $N^T$ nor $N^\perp$ individually commute with $\partial_a$'s. In fact $[N^T,\partial_a]=-[N^\perp,\partial_a]$ and $[N^T,\partial_a]^\perp=-[N^\perp,\partial_a]^\perp=0$. We will rewrite $K(\partial_a,\partial_b)$ as $\nabla_{\partial_a}\partial_b-(\nabla_{\partial_a}\partial_b)^T$.
 \begin{gather}
 h^{ab}\nabla_N\nabla_{\partial_a}\partial_b=R(N,\partial_a)\partial_b+\nabla_{\partial_a}\nabla_{\partial_b}N
 \end{gather}
 The derivative of the tangent part gives,
 \begin{gather}
 h^{ab}\nabla_N(g(\nabla_{\partial_a}\partial_b,\partial_c)h^{cd}\partial_d)=\bigg(R(N,\partial_a)\partial_b+\nabla_{\partial_a}\nabla_{\partial_b}N\bigg)^T+g(H,(\nabla_{\partial_c}N)^\perp )h^{cd}\partial_d+(\nabla_{(\nabla_{\partial_a}\partial_b)^T}N)^\perp
 \end{gather}
 The $\nabla_N h^{ab} K(\partial_a,\partial_b)$ term gives,
 \begin{equation}
  \nabla_N h^{ab} K(\partial_a,\partial_b)=-h^{cd}K(\partial_a,\partial_c)\bigg[g(\nabla_{\partial_b}N^\perp,\partial_d)+g(\partial_b,\nabla_{\partial_d}N^\perp)\bigg]-2h^{ae}h^{bf}\bigg(g(\partial_e,D_{\partial_f}N^T)\bigg)K(\partial_a,\partial_b)
 \end{equation}
 Meanwhile $\delta_gH= \accentset{(1)}{C}(\partial_a,\partial_b))^{\perp}-h^{ac}h^{bd}K(\partial_a,\partial_b)\accentset{(1)}{P}(\partial_c,\partial_d)$. Using the identities in section(\ref{N&C}) and using the above expressions along with the results of appendix (\ref{TVH}), one can write $\delta_N H$ as, 
 \begin{gather}
 h^{ab}\Bigg({B}_{ab}-{D}_{ab}+C(\partial_a,\partial_b)+S_{ab}-h^{cd}K(\partial_a,\partial_c)P(\partial_d,\partial_b)\Bigg)^\perp-g(H,(\nabla_{\partial_c}N)^\perp )h^{cd}\partial_d+\nabla_{N^T}H,\label{FV}
 \end{gather}
 where the quantities ${B}_{ab},{D}_{ab}$ and $S_{ab}$ are given by
 \begin{eqnarray*}
 &{B}_{ab}:=R(N^\perp,\partial_a)\partial_b+\nabla_{\partial_a}\nabla_{\partial_b}N^\perp,~~~~~ 
 {D}_{ab}:=\nabla_{(\nabla_{\partial_a}\partial_b)^T}N^\perp,\\ &{S}_{ab}:=-h^{cd}K(\partial_a,\partial_c)\bigg[g(\nabla_{\partial_b}N^\perp,\partial_d)+g(\partial_b,\nabla_{\partial_d}N^\perp)\bigg]
 \end{eqnarray*}
 For convenience we will define one more quantity $L_{ab}$, which will be useful later and is given as 
 \begin{equation*}
 L_{ab}=\left({B}_{ab}-{D}_{ab}+C(\partial_a,\partial_b)+S_{ab}-h^{cd}K(\partial_a,\partial_c)P(\partial_d,\partial_b)\right)
 \end{equation*}
 The appearance of the $\nabla_{N^T}H$ might have have looked obvious but is not quite so. Due the fact that neither $N^T$ nor $N^\perp$ commute with the $\partial_a$'s there could have been terms, depending on the commutators, surviving. It however turns out that the commutator terms do not contribute (appendix \ref{TVH}). Since $H=0$ everywhere on $\mathcal S$ it follows that $\nabla_{N^T}H$ =0. The resulting form can then be shown to reproduce the correct form of the Jacobi equation once the minimal surface condition $H=0$ is imposed. 
 
 \begin{gather}
 \mathcal{L}N^{\perp}=-C^\perp+\tilde{H}.\label{IJE}
 \end{gather}
 where
 \begin{gather}
 \mathcal{L}(N^{\perp})=\Delta^{\perp}N^{\perp}+Ric(N^{\perp})+A(N^{\perp}),\label{devi2}
 \end{gather}
 where we have defined  $\Delta^{\perp}N^{\perp}$ to be the Laplacian on the normal bundle, given by $h^{ab}\Bigl(\nabla^{\perp}_{\partial_a}\nabla^{\perp}_{\partial_b}N^{\perp}-\nabla^{\perp}_{(\nabla_{\partial_a}\partial_b)^T}N^{\perp}\Bigr)$, $g(R(N^{\perp},\partial_a)\partial_b,N^{\perp})$ has been denoted by $Ric(N)$. $A(N)=h^{ab}K(\partial_a,W_{N}(\partial_b))$ is the Simon's operator. Whereas  $C^\perp$ is defined as $C^\perp=h^{ab}\accentset{(1)}{C}(\partial_a,\partial_b)^\perp$ and $\tilde{H}=\accentset{(1)}{P}^{ab}K(\partial_{a},\partial_{b})$.
 Thus identifying the Jacobi/Stability operator $(\mathcal L)$ for minimal surfaces. This equation was derived in \cite{Ghosh:2017ygi} in the context of the co-dimension two entangling surface, but will be applicable our case as well where the complexity surface is also minimal. In fact this equation is applicable to any embedded minimal submanifold ( $\mathscr S$ say ) and the solutions of the above equation $N^\perp$ maps a minimal submanifold ($\mathscr S$) to another submanifold ($\mathscr S'$ ) which is minimal in a perturbed geometry. Hence, the superscript $\perp$ here implies components normal to $\mathscr S$ and $T$ implies components tangential to $\mathscr S$.

 The form eq. (\ref{FV}) will be taken to perform the second variation. The reason for taking this and not the known form is the following. Inside a derivative one cannot assume $N^\perp$ to be normal, hence the term $S_{ab}$ cannot be written in terms of the extrinsic curvature inside a derivative. If one is dealing with a family of co-dimension two surfaces then taking $N^\perp$ inside a derivative amounts to making an assumption that $N^\perp$ is normal throughout the foliation. The simple analogue of this statement is the following. $g(N^\perp,\partial_a)$ is equal to zero, but its normal derivative $\nabla_{N^\perp}g(N^\perp,\partial_a)$ may not be.
 
 \subsubsection{Second order perturbation equations}\label{soperte}
 The second order equation is obtained by equation the second total variation of $H$ to zero, i.e,
 \begin{equation}
 (\mathcal L_N H+\delta_g)(\mathcal L_N H+\delta_g) H=0
 \end{equation}
 One can show that this is nothing but,
 \begin{eqnarray}
 (\nabla_N H+\delta g)(\nabla_N H+\delta g) H -R(N,H)N-\nabla_{[N,H]}-\nabla_{\delta_g H}N-\nabla_H\delta_g N=0,
 \end{eqnarray}
 where the condition $\mathcal L_N H+\delta_g N=0$ condition has been used for the back-ground. It is evident from the the above expression that on further this condition along with the condition $H=0$ on the back-ground yields the equation,
 \begin{equation}
 (\nabla_N H+\delta g)(\nabla_N H+\delta g) H=0
 \end{equation}
 Note that the equation $H=0$ holds everywhere on $\mathcal S$. Therefore $\nabla_{N^T}\nabla_{N^T} H=0$. Further, since the inhomogeneous Jacobi equation holds everywhere on $\mathcal S$, it's tangent derivative is also zero. Therefore while finding the second variation one only needs to consider the normal component $N^\perp$ of the deviation vector and not the tangent part. The final expression that needs to be calculated is therefore $(\nabla_{N^\perp}+\delta_g)\bigg({L}_{ab}^\perp\bigg)$. Further note that $\delta_g N$ has been taken to be zero. This is because had this been taken to be non-zero one would have got an equation for $\nabla_N^{\perp}N^\perp+\delta_g N^\perp$ rather than $\nabla_{N^\perp}N^\perp$. Every expression including the area variations would then have contained  $\nabla_N^{\perp}N^\perp+\delta_g N^\perp$ rather than $\nabla_{N^\perp}N^\perp$. Hence $\delta_g N$ is just a gauge redundancy and can be set to zero. Everything put together given the following equation,

\begin{gather}
h^{ab}\bigg(\nabla_{\partial_a}\nabla_{\partial_b}J^\perp-\nabla_{(\nabla_{\partial_a}\partial_b)^T}J^\perp+R(J^\perp,\partial_a)\partial_b+2h^{cd}K(\partial_c,\partial_a)g(J^\perp,K(\partial_b,\partial_d))\bigg)^\perp\notag\\
+h^{ab}h^{cd}L_{ac}^\perp \bigg[4g(N,K(\partial_b,\partial_d))-2P(\partial_b,\partial_d)\bigg]-2h^{ab}\nabla_{(L_{ab})^T}N\notag\\
-h^{ab}h^{cd}K(\partial_c,\partial_d)\bigg[2g(R(N,\partial_b)N,\partial_d)+2g(\nabla_b N,\nabla_d N)\bigg]\notag\\
h^{ab}\bigg[4R(N,\partial_a)\nabla_{\partial_b}N+(\nabla_{\partial_a}R)(N,\partial_b)N+(\nabla_N R)(N,\partial_a)\partial_b\bigg]\notag\\
h^{ab}\bigg[2(\nabla_N C)(\partial_a,\partial_b)+4C(\nabla_{\partial_a}N,\partial_b)\bigg]-4h^{ab}h^{cd}K(\partial_a,\partial_c)\bigg[g(C(N,\partial_b),\partial_d)+P(\nabla_{\partial_b} N,\partial_d)\bigg]\notag\\
h^{ab}\accentset{(2)}{C}(\partial_a,\partial_b)^\perp-2P(C(\partial_a,\partial_b)^\perp)-h^{ab}h^{cd}\accentset{(2)}{P}(\partial_b,\partial_d)K(\partial_a,\partial_c)=0.\label{2IJE}
\end{gather}
In order to find out the new minimal surface in $\mathcal M'$ one needs to find out what $N$ on $\mathcal S$ is, which is provided by the solution to eq. (\ref{IJE}) and all derivatives of $N$ along $N$ itself, on $\mathcal S$. It is precisely this that these perturbation equations provide. The second order perturbation equation eq. (\ref{2IJE}) for example is providing what $(\nabla_NN)^\perp$ is. Further higher order equations can in general be obtained, but is not useful to our discussion.

 \subsection{The perturbation scheme and solution procedure}\label{pscheme}
 The scheme for calculating the expressions eq. (\ref{c}, \ref{sc}) is the following. The first term in the first order variation eq. (\ref{c}) vanishes by equations of motion. Hence, one has to calculate the second term and the boundary term. The second term can easily be calculated as the embedding of $\mathscr S$ and the perturbation $\accentset{(1)}{P}(\partial_a,\partial_b)$ are known. Calculating the boundary term however requires more information. Mathematically one requires $N\mid_{\partial\mathscr S}$. This  information is provided by how the boundary $\partial \mathscr S$ deforms. Let the vector field $\eta$ characterise this deviation. Therefore one has $N\mid_{\mathscr S}=\eta$. The Jacobi equation on $\mathscr S$ is an expression for $N^\perp$. The boundary conditions for solving this equation is of the Dirichlet type i.e $N^\perp\mid_\mathcal S$ is given. It is precisely this boundary condition that is provided by $\eta^\perp$, while $\eta^T$ can be used for calculating the change in volume using eq. (\ref{c}). At second order one has to evaluate the terms in eq. (\ref{sc}). All the terms except the last one in this expression can be obtained using the expressions for $N^\perp$, which are provided by solution of the first order IJE for $\mathscr S$. The last term requires a discussion of the second order IJE. The second order IJE provides an expression for $(\nabla_N N)^\perp$ the boundary condition for which is fed from $\nabla_\eta\eta$ i.e $(\nabla_N N)^\perp\mid_{\partial\mathscr S}=(\nabla_\eta\eta)^\perp$ while $(\nabla_\eta\eta)^T$  is used in eq. (\ref{sc}).
 
 Let us now discuss the solution procedure for the special case we will be dealing with. Note $\eta$ is completely uspecified. Now, consider the special case where $\partial\mathscr S$ is also a minimal surface, but ofcourse, of one dimesion less than $\mathscr S$. Assume that under the perturbation $\mathscr S$ deform to a minimal surface $\mathscr S'$ such that $\partial \mathscr S$ also deforms to a minimal surface $\partial \mathscr S'$. Then it is evident that $\eta$ is a solution of an IJE but now for $\partial\mathscr S$ rather than of $\mathscr S$ i.e,
 
 \begin{gather}
 \mathcal{L}N^{\perp}=-C^\perp+\tilde{H}.\label{IJE2}
 \end{gather}
 where
 \begin{gather}
 \mathcal{L}(N^{\perp})=\Delta^{\perp}N^{\perp}+Ric(N^{\perp})+A(N^{\perp}),\label{devi}
 \end{gather}
 where we have defined  $\Delta^{\perp}N^{\perp}$ to be the Laplacian on the normal bundle of $\partial\mathscr S$, given by $\gamma^{AB}\Bigl(\nabla^{\perp}_{\partial_A}\nabla^{\perp}_{\partial_B}\eta^{\perp}-\nabla^{\perp}_{(\nabla_{\partial_A}\partial_B)^T}\eta^{\perp}\Bigr)$, $g(R(\eta^{\perp},\partial_A)\partial_B,\eta^{\perp})$ has been denoted by $Ric(\eta)$. $A(\eta)=\gamma^{AB}K(\partial_A,W_{\eta}(\partial_B))$ is the Simon's operator. Whereas  $C^\perp$ is defined as $C^\perp=\gamma^{AB}\accentset{(1)}{C}(\partial_A,\partial_B)^\perp$ and $\tilde{H}=\accentset{(1)}{P}^{AB}K(\partial_{A},\partial_{B})$. All these quantities now refer to $\partial\mathscr S$ as opposed to $\mathscr S$ and $\perp$ now refers to perpendicular to $\partial\mathscr S$ rather than $\mathscr S$. Similarly $\nabla_\eta\eta$ is obtained from the second order IJE eq. (\ref{2IJE}) but now written for $\partial\mathscr S$ rather than for $\mathscr S$ by making the same changes as done for eq. (\ref{IJE2}).

\section{Perturbative change of subregion complexity}\label{FOV}
Given an asymptotically $AdS$ spacetime and assuming that it is dual to a CFT, the holographic entanglement entropy of a subregion $A$ on a fixed time slice in the boundary CFT is given according to the Ryu-Takayanagi proposal \cite{Ryu:2006ef}, by the area of an extremal (minimal surface in pseudo-Riemannian geometry \footnote{From now on we will use extremal and minimal interchangeably.}) co-dimension two surface ($\mathcal S$) in the bulk, which is homologous to the boundary $\partial A$ of the subregion $A$. 
The holographic complexity is given by the volume ($V$) of an extremal co-dimension one minimal surface ($\Sigma$) with $\mathcal S\cup A$ as its boundary \cite{Stanford:2014jda}. 
\begin{gather}
C_V =\frac{V}{8\pi R G}\label{new1}
 \end{gather}

For a static geometry the $t=constant$ slices ($T$ say) are maximal and hence both $\mathcal S$ and $\Sigma$ lie on $T$ and therefore all the calculations can be done on the $T$'s. For non-static or non stationary bulk spacetimes this may not be the case. In such situations one uses the covariant Hubeny-Rangamani-Takayanagi (HRT)\cite{Hubeny:2007xt} proposal to calculate the holographic entanglement entropy. It is known from the variational approach that the contribution from the changes in embedding does not appear at the first order of metric perturbation \cite{Nozaki:2013vta,Bhattacharya:2013bna}. Hence one can work with the same fixed time slice in non static spacetimes for calculating HEE upto first order. In this section we develop a similar variational scheme to handle this situation for the volume of subregions. Unlike area the first variation of the volume functional incorporates the changes in the embeddings. However in the next section we will see that the information of the deviation from the $t=constant$ time slice does not appear in the first variation. Hence one can still work with the fixed time slice for the volume change in non static backgrounds.

Let us denote by $\mathcal S'$ and $\Sigma '$ the entangling surface and the complexity surface in a perturbed geometry. We will be coordinatizing both the primed and the unprimed surfaces by the same coordinate system. $\Sigma$ will now play the role of $\mathscr S$ and $S\cup A$ the role of $\partial\mathscr S$ of section \ref{VGM}. The area of $\mathscr S$ is nothing but the volume ($V$) of the spacelike surface $\Sigma$. $\Sigma '$ can then be characterised by a deviation vector $N$ between $\Sigma$ and $\Sigma'$. Let the restriction of $N$ to $\mathcal S\cup A$ be denoted by $\eta$. Since, the boundary subsytem is not deformed it is evident that $\eta\mid_{A}=0$. Further $\eta$ must be such that it maps the minimal surface $\mathcal S$ to the minimal surface $S'$ in the perturbed metric. The change in volume of the complexity surface is then given by,
\begin{eqnarray}\label{firstvariation}
&&\delta_{N} V=-\int_\Sigma g\left(N,K^{(\Sigma\hookrightarrow\mathcal M})(\partial_a,\partial_b)\right)h^{ab}\sqrt{h}~d^{d}\tau+\frac{1}{2}\int_\Sigma \accentset{(1)}{P}(\partial_a,\partial_b)h^{ab}\sqrt{h}~d^{d}\tau\notag\nn
&&-\int_{\mathcal S ~\cup ~ A} g(\hat r, N)\sqrt{\gamma}~d^{d-1}\chi,
\end{eqnarray}
where $\tau^a$ are the coordinates intrinsic to $\Sigma$, $\chi^A$ are the coordinates intrinsic to $\mathcal S$, $K^{(\Sigma\hookrightarrow\mathcal M})(\partial_a,\partial_b)$ is the extrinsic curvature of $\Sigma$ in $\mathcal M$ and $\hat r$ is the inward directed co-normal of $\mathcal S$ w.r.t $\Sigma$. Note that the first term does not contribute since $\Sigma$ is extremal. Moreover $N$, in the last term, can be replaced by its restriction $\eta$ to $\mathcal S$. The component of $N$ perpendicular to $\Sigma$, in the above expression, is obtained by solving the inhomogeneous Jacobi equation (IJE) eq. (\ref{IJE}).

 To make things clear we will put another subscript to $\perp$ and $T$ ($\perp_{\Sigma},~T_\Sigma$ for example) to indicate the surface to which the the components are perpendicular and tangential. As has been emphasized in section \ref{pscheme}, this eq. (\ref{IJE}) does not determine the tangential component $N^{T_\Sigma}$. $N^{T_\Sigma}$ which is known only at the boundary $\mathcal S\cup A$ and is obtained by how $\mathcal S$ and the subregion $A$ deforms. It can also be shown that the volume change does not depend on how $N^{T_{\Sigma}}$ is extended to $\Sigma$. It is clear from the situation at hand that $A$ does not deform and the deformation is non-zero only on $\mathcal S$. However note that both $\mathcal S$ and $\mathcal S'$ are co-dimension two minimal surfaces in $\mathcal M$ ans $\mathcal M'$. Hence $\eta:=N\vert_{\mathcal S}$ takes a co-dimension two minimal surface in $\mathcal M$ to one in $\mathcal M'$. Therefore, $\eta^{\perp_\mathcal S}$ must satisfy an inhomogeneous Jacobi equation (\ref{IJE2}) for a co-dimension two minimal surface $\mathcal S$. Note that $\eta$ can be taken to be completely perpendicular to $\mathcal S$. This is because the boundary of $\mathcal S$ which is nothing but $\partial A$, does not deform under the perturbation as the boundary subregion is kept intact. Since the normal bundle of $\mathcal S$ is two dimensional, there are essentially two independent components of $\eta$. One of the components will be perpendicular to both $\Sigma$ and $\mathcal S$ and the other would be perpendicular to $\mathcal S$ but tangent to $\Sigma$. It is precisely the second component that is along the co-normal $\hat r$. Hence, $\big(\eta\big)^{\perp_\Sigma}=\big(\eta^{\perp_\mathcal S}\big)^{\perp_\Sigma}:=N^{\perp_{\Sigma}}\vert_{\mathcal S}$ determines the boundary condition for eq. (\ref{IJE}) applied to $\Sigma$ whereas $\eta^{T_\Sigma}=\big(\eta^{\perp_\mathcal S}\big)^{T_\Sigma}$ determines the change in volume in eq. (\ref{firstvariation}). The solutions $\eta$ for $\mathcal S$ was obtained for certain cases in \cite{Ghosh:2016fop,Ghosh:2017ygi}. We will essentially be using these solutions to determine the change in volume from eq. (\ref{firstvariation}). 

In the next few subsections we will be evaluating the expression eq. (\ref{c}) for several examples, for which the deviation vector has been calculated in \cite{Ghosh:2016fop,Ghosh:2017ygi}. First we will consider the $2+1$ dimensional case in which we will be considering rotating BTZ as a perturbation over $AdS_3$ as an example. Next, for the $3+1$ dimensional case we will be considering boosted black brane like perturbations over $AdS_4$. 

\subsection{2+1 dimensional case}\label{2+1}
Consider the $AdS_3$ metric in the Poincar\'e patch as the unperturbed background and consider the planar rotating BTZ as a perturbation over it. The $AdS_3$, metric is given by
\beq
ds^2=\frac{-dt^2+dz^2+dx^2}{z^2}
\eeq
  $\mathcal S$ in this case is given by a spacelike geodesic in $AdS_3$. The equation for such a geodesic, which is homologous to an interval on a spacelike slice of the boundary is given by,
\beq
z(\tau)={z^{(0)}_{*}}\sech(\tau),~~~x(\tau)={z^{(0)}_{*}}\tanh{(\tau)},
\eeq.
$z_*^{(0)}$ is the $AdS$ turning point. The parameter~$\tau~\epsilon~~(-\infty,\infty)$ covers the full geodesic.  Since the geodesic is affinely parametrised the metric on $\mathcal S$ is just identity. The metric on $\Sigma$
can be conveniently written in a co-ordinate system that includes $\tau$ and another co-ordinate $c$ (say). To understand how $c$ is defined note that $z_*^{(0)}$ is a constant on a particular geodesic. However this can be lifted to the coordinate $c$ (say) so that $c,~\tau$ cover the region bounded by $\mathcal S$ and $A$ on a $t=constant$ slice ($\Sigma$) \footnote{This follows from the nesting property of entanglement entropy \cite{Wall:2012uf}. In other words as one reduces the subregion size the bulk geodesics do not intersect each and therefore the turning point behaves monotonically and qualifies for a good coordinate.}. In other words we are foliating this region in such a way that, each leaf is a spacelike geodesic with a particular turning point, given by the value of $c$ on that leaf. The metric $h$ on $\Sigma$ is then given by,

\begin{equation}
h_{ab}~d\tau^ad\tau^b=\frac{dc^2+\sech(\tau)^2~d\tau^2}{c^2\sech^2{\tau}}.
\end{equation}

The deviation vector $\eta$ can be calculated by solving an inhomogenous Jacobi equation for the spacelike geodesic in $AdS_3$. Given the tetrads that are parallely propagated along the geodesic,
\beq
e^\mu_0=(z,0,0),~~e^\mu_1=\left(0,\pm z\sqrt{1-\left(z\over{z^{(0)}_{*}}\right)^2},{z^2\over{z^{(0)}_{*}}}\right),~~e^\mu_2=\left(0,{z^2\over{z^{(0)}_{*}}},\mp z\sqrt{1-\left(z\over{z^{(0)}_{*}}\right)^2}\right),
\eeq
$\eta$ can be written as $\eta^I~e_I^\mu$, with,
\beq
\eta^0&=&\frac{1}{3}b {z^{(0)}_{*}}^2 \tanh(\tau)\sech(\tau)\\
\eta^2&=&-\frac{1}{6} a {z^{(0)}_{*}}^2\cosh(2\tau)\sech(\tau)^3
\eeq
Note that the co-normal $\hat r$ in eq. (\ref{firstvariation}) is nothing but $-e^\mu_2$. Hence the last integral in eq. (\ref{firstvariation}) reduces to integrating $\eta^2$ over the spacelike geodesic. The integral evaluates to,

\beq
\int g(\hat r, \eta)\sqrt{h}~d^{d-1}\tau=\frac{\pi}{4} a {z^{(0)}_{*}}^2
\eeq
To evaluate the first term, we will use the $\tau,~c$ coordinatisation.  Then calculating the first term reduces to calculation the integral.
\beq
a\int_{-\infty}^{\infty}\int_0^{z^{(0)}_{*}}c\sech^3(\tau) dc~d\tau=\frac{\pi}{4} a {z^{(0)}_{*}}^2
\eeq
The total change in volume can then be calculated as,
\begin{eqnarray}
\Delta V=\frac{\pi}{8}a {z^{(0)}_{*}}^2
\end{eqnarray}

The only question remaining is whether the inhomogenous Jacobi equation for the co-dimension one minimal surface admit a solution with the boundary conditions given by solutions of the co-dimension two inhomogeneous Jacobi equation. To see it, let us write down the equation first. The normalized normal one form of the $t=constant$ slices are given by $\frac{dt}{z}$, and the normal vector is given by $z\partial_t$. Let us denote it by $\hat{n}$. Since this is the only normal and is also normalized, it follows that the $\nabla_a^\perp\hat{n}=0$. Now, assume that the solution is of the form $\alpha \hat{n}$. The codimension one inhomogeneous Jacobi equation, for rotating BTZ like perturabtions reduces to,

\begin{equation}\label{1DJ}
z^2\left(\partial_x^2 \alpha+\partial_z^2 \alpha\right)-2\alpha=0.
\end{equation}
Note that the inhomogeneous term is in fact zero for this case. This can be checked from the explicit form of $\accentset{(1)}{C}^\mu~_{\nu\rho}$ given in appendix (\ref{FG3}).
To facilitate imposing boundary conditions on the entangling surface we will work in coordinates $c, ~\tau$ as explained before.

 \begin{eqnarray}
 \sech^2\tau(y^2\partial_y^2\alpha+y\partial_y\alpha)+\partial_\tau^2\alpha+\tanh\tau\partial_\tau\alpha-2\alpha=0
 \end{eqnarray}
 Boundary conditions,
 \begin{eqnarray}
 \alpha(y_0,\tau)=\frac{b}{3}y_0^2\tanh\tau\sech\tau,~~~\alpha(y,\tau=\infty)=0,~~~\alpha(y,\tau=-\infty)=0
 \end{eqnarray}
 \begin{figure}
 	\includegraphics[width=8cm]{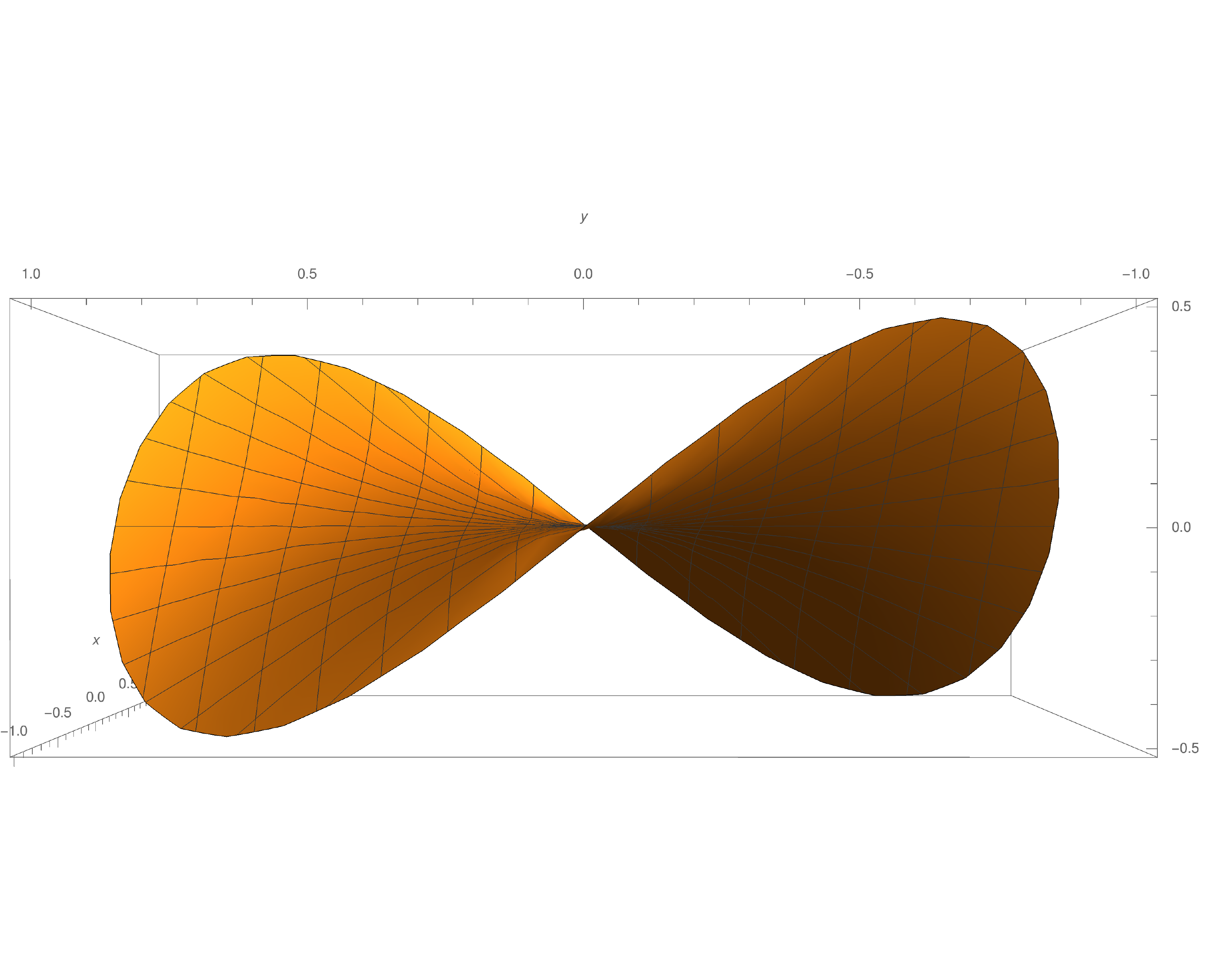}~~~
 	\includegraphics[width=8cm]{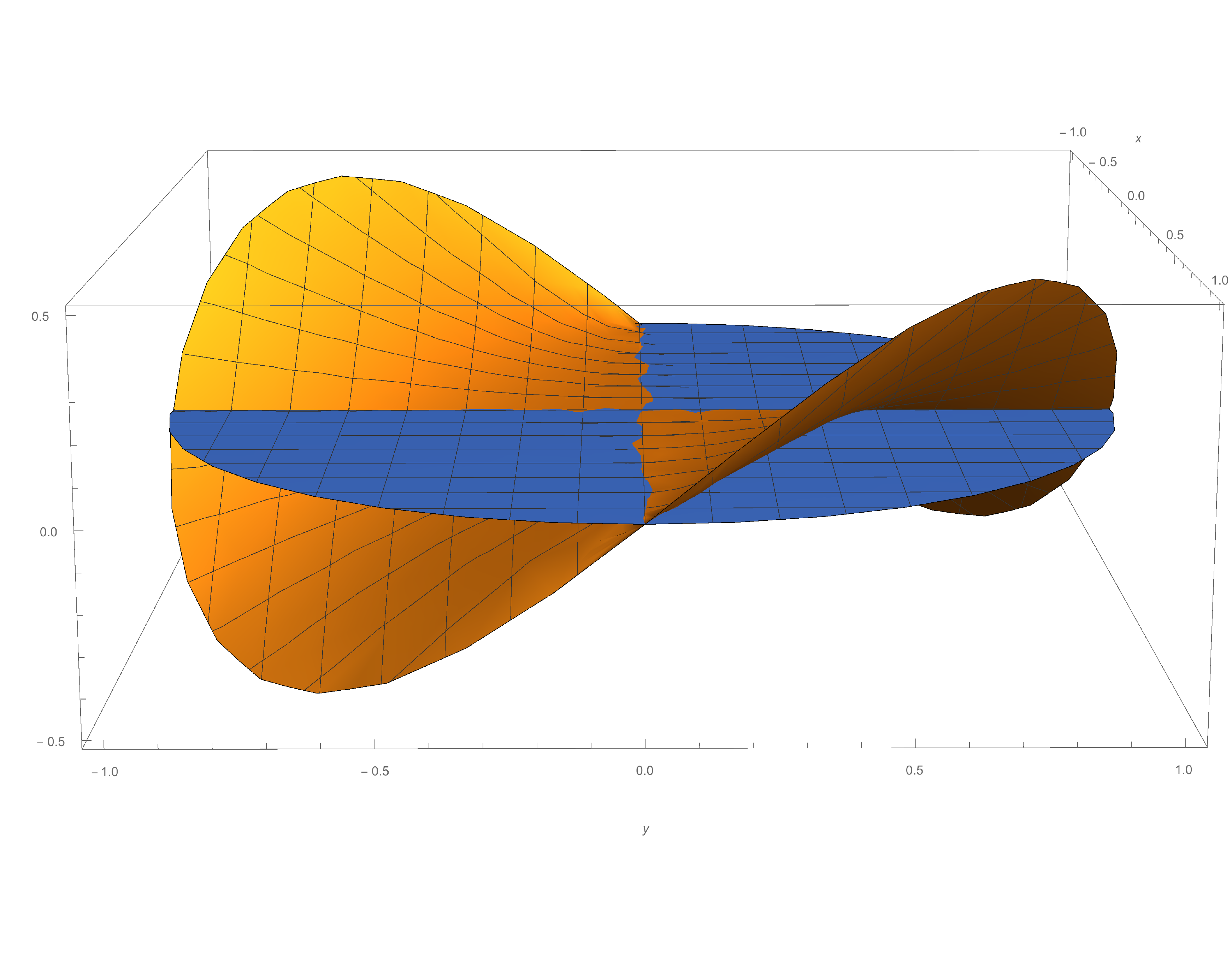}
 	\caption{Solution of the co-dimension one minimal surface equation with boundary conditions given by the co-dimension two minimal surface equation. Note that the solution has been obtained by solving the equation on a disc rather than on a half disc as required. The first figure confirms that the boundary condition at $z=0$ is indeed statisfied. The second figure shows the normal deviation of the minimal surface from the unperturbed one.}
 	\label{1DJFIG}
 \end{figure}
 Unfortunately this equation does not admit a solution by separation of variables. We therefore try to solve eq. (\ref{1DJ}) on the disc $x^2+z^2=1$ with the first boundary condition only (The actual solution is however required on the half disc with the full set of boundary condition). If the solution satisfies the second and the third boundary condition then we can conclude that a solution on the half disc with full set of boundary conditions does exist. The plots of the solutions are given in fig. (\ref{1DJFIG}). From the figure it is clear that this indeed is the case.

\subsection{3+1 dimensional case}
For the $3+1$ dimensional case we will consider the Boosted AdS black brane like perturbations over $AdS_4$. The metric and the asymptotic expansion of the metric is given in appendix \ref{FG4}. The choice of the back ground metric is therefore the $AdS_4$ metric in the Poincar\'e patch. The AdS length is set to unity and the metric is given by,
 \begin{gather}
  ds^{2}= {-dt^2+dx^2+dy^2+dz^2\over z^2}
 \end{gather}
 We will now consider a circular disk like subregion and calculate the change in holographic complexity from pure the AdS value.
 
 \subsubsection{Circular disk subregion}
For the case of a circular boundary subregion of radius $\mathscr R$, the minimal surface in the $AdS_{d+1}$ is a $d-1$ dimensional hypersphere. Such a surface can be described by the following embedding functions, which are just a parametric form  of the equations obtained in \cite{Bhattacharya:2013bna, Fonda:2014cca},
 \begin{gather}
 x=\mathscr R\sin{\theta}\cos{\phi}+X,~~y=\mathscr R\sin{\theta}\sin{\phi}+Y,~~z=\mathscr R\cos{\theta},~~t=constant.\label{embedsphere}
 \end{gather}
The coordinates $\theta, \phi$ have ranges, $0\le\theta\le\frac{\pi}{2}$ and $0\le\phi<2\pi$. It is also clear from eq.(\ref{embedsphere}) that the half sphere intersects the $AdS_4$ boundary at $\theta=\frac{\pi}{2}$. The intrinsic metric on $\mathcal S$ is then given by,
\begin{gather}
\gamma_{AB}~d\chi^Ad\chi^B={d\theta^2 +\sin^2{\theta}~ d\phi^2\over\cos^2{\theta}},
\end{gather}
while the metric on $\Sigma$ in $\mathscr R,~\theta,~\phi$ coordinates is given by,
\begin{equation}
h_{ab}~d\tau^ad\tau^b=\frac{d\mathscr R^2+\mathscr R^2(d\theta^2+\sin^2\theta~d\phi^2)}{\mathscr R^2\cos^2\theta}
\end{equation}

We will set up a local orthonormal basis on this surface to facilitate our calculations. Since the surface is space-like, the vectors
\begin{gather}
e_{2}=\cos{\theta}\partial_{\theta},~~e_{3}=\cot{\theta}\partial_{\phi}.
\end{gather}
provide us with two space like bases. The other two, spanning the normal bundle will provide us with the other two basis vectors. As a matter of convention we mark the time like normal as $e_0$ and the space like normal as $e_1$.
\begin{gather}
e_{0}=z\partial_{t},~~e_{1}={z(x-X)\over \mathscr R}\partial_{x}+{z(y-Y)\over \mathscr R}\partial_{y}+{z^2\over \mathscr R}\partial_{z}
\end{gather}

The solution to the co-dimesion two inhomogeneous Jacobi equation has been obtained in \cite{Ghosh:2017ygi} and can be represented as the components of the deviation vector in the directions normal to the co-dimension two surface eq. (\ref{embedsphere}) \footnote{The solution given in \cite{Ghosh:2017ygi} has an error. The authors missed a factor of $\frac{1}{2}$ in the first term of the inhomogeneous part of eq. (42) of \cite{Ghosh:2017ygi}.},
\begin{gather}
\eta^1=\frac{\mathcal R ^3}{96}  \left(3 \beta ^2 \gamma ^2+2\right) \cos ^2\theta \big(\cos 2 \theta-5\big)-\frac{\mathcal R ^3}{64} \cos 2 \phi \bigg(\beta ^2 \gamma ^2 \sin ^22 \theta\bigg)
\end{gather}
\begin{gather}
\eta^0=-\frac{1}{4} \beta \gamma ^2 \mathcal R^3 \sin \theta \cos ^2\theta\cos{\phi}
\end{gather}
The third term in eq. (\ref{firstvariation}) is then just the integral of $-N^1$ over $\mathcal S$ and yields,
\beq
\int g(\hat r, \eta)\sqrt{h}~d^{d-1}\tau =\frac{\pi}{9z_0^3}  \mathscr R^3 \left(3 \beta ^2 \gamma ^2+2\right) 
\eeq
The second term in eq. (\ref{firstvariation}), can be calculated by lifting $\mathscr R$ to a coordinate $r$. The expression then reduces to,
\beq
\frac{3 \beta ^2 \gamma ^2+2}{z_0^3}\int_0^{\mathscr R}\int_0^{\frac{\pi}{2}}\int_0^{2\pi}r^2\sin(\theta)dr~d\theta~d\phi=\frac{2\pi}{9z_0^3}\mathscr R^3(3 \beta ^2 \gamma ^2+2)
\eeq
The total change in complexity is then given by,
\beq
\Delta C=0
\eeq
Thus the first order correction to the holographic complexity vanishes for a spherical entangling surface. This is consistent with the result for AdS black brane ($\beta\to 0$) results obtained in \cite{Alishahiha:2015rta,Ben-Ami:2016qex}. However, the finite $\beta$ case has not been dealt with before. Hence, we can conclusively say that even for a boosted black brane like perturbations over $AdS$ the first order change of subregion comlexity is zero fro a spherical subregion.

 \subsubsection{Thin Strip subregion}
 Let us now consider a two dimensional strip like subregion on the $AdS_4$ boundary, given by the region $[-L,L]\times[-\frac{l}{2},\frac{l}{2}]$ of the $x-y$ plane, where $L\gg l$. The minimal entangling surface in this case, given in \cite {Fonda:2014cca}, can be characterized by the following embedding functions,
 \begin{gather}
 x=\lambda,~~y(\theta)=-z_* E\left({\pi-2\theta\over 4}\left.\right|2\right),~~z(\theta)=z_*\sqrt{\sin{\theta}},\label{embstrip}
\end{gather}
where $z_{*}$ is the turning point of the minimal surface in $AdS_4$ and $E(\alpha,\beta)$ is the incomplete elliptic integral of the second kind. The range of the coordinates  are $0\le\theta\le\pi$ and $-L\le\lambda\le L$. The intrinsic  metric of $\mathcal S$ therefore takes the form
\begin{gather}
\gamma_{AB}~d\chi^Ad\chi^B={{z_*}^2d\theta^2+4\sin{\theta}d\lambda^2\over 4 z_*^2\sin{\theta}^2}.
\end{gather}
 Further, the turning point $z_{*}$ can be written in terms of the width $l$ of the subregion as $z_*=\frac{\Gamma(\frac{1}{4})l}{2\sqrt{\pi}\Gamma(\frac{3}{4})}$.The metric on $\Sigma$ can be written in a co-ordinate $c,~\lambda,~\theta$, where $c$ is obtained by elevating the constant $z_*$ to a coordinate. We again choose a tetrad adapted to the $\mathcal S$ as,
\begin{gather}
 e_2=2\sin{\theta}\partial_{\theta},~~e_3=z_*\sqrt{\sin{\theta}}\partial_{\lambda},~~ e_1=z(\sin{\theta}\partial_z-\cos{\theta}\partial_y),~~e_0=z\partial_t
\end{gather}

\subsubsection*{\underline{ Case A: Strip along `x' boost along `x'}}\label{SAXBAX}
The component of the solution that will be important for our case is $\eta^1$, which is given in \cite{Ghosh:2017ygi},
\begin{gather}
\eta^1=\frac{C_1\cos(\theta)}{\sqrt{\sin(\theta)}}+C_2\sin{\theta}~~_2F_1\bigg(\frac{1}{4}, 1; \frac{1}{2}, \cos^2\theta\bigg)\notag\\
-\frac{\cos(\theta)}{\sqrt{\sin(\theta)}}\int_{}^{\theta}\left[\frac{1}{4}\left(3D+\frac{3C}{2}\right)z_*^3\left(\sin{\theta}\right)^{\frac{1}{2}}-\frac{7}{8}Dz_*^3\left(\sin{\theta}\right)^{\frac{5}{2}}\right](\sin{\theta'})^{\frac{3}{2}}~~_2F_1\bigg(\frac{1}{4}, 1; \frac{1}{2}, \cos^2\theta'\bigg)d\theta'\notag\\
+\sin{\theta}~~_2F_1\bigg(\frac{1}{4}, 1; \frac{1}{2}, \cos^2\theta\bigg)\int_{}^{\theta}\left[\frac{1}{4}\left(3D+\frac{3C}{2}\right)z_*^3\left(\sin{\theta}\right)^{\frac{1}{2}}-\frac{7}{8}Dz_*^3\left(\sin{\theta}\right)^{\frac{5}{2}}\right]\cos(\theta')d\theta'\label{SAXBAXSOL}
\end{gather}
the constants of integration being given by $C_1=\frac{\pi z_*^3}{16} \left(2 C+D\right)$ and $C_2=-\frac{\Gamma \left(\frac{1}{4}\right)^2 z_*^3\left(2C+D\right)}{16 \sqrt{2 \pi }}$ and constants $C,D$ are given by $C=\left({1\over 3}+\beta^2\gamma^2\right){1\over z_0^3},~D={1\over 3}{1\over z_0^3}$ (appendix \ref{FG4}). The integral over $\Sigma$ can be evaluated in $c,~\lambda,~\theta$ coordinate. The integral to be evaluated then reduces to,

\begin{gather}
\frac{(\frac{2}{3}+\beta^2\gamma^2)}{z_0^3}\int_{-L}^L\int_0^{z_*}\int_0^\pi c \bigg(\frac{1}{2} \sqrt{\sin\theta} \sqrt{1-2 \sin ^2\left(\frac{1}{4} (\pi -2 \theta)\right)}+\frac{\cos\theta ~E\left(\left.\frac{1}{4} (\pi -2 \theta)\right|2\right)}{2 \sqrt{\sin\theta}}\bigg)dc~d\theta\notag\\
=\frac{2Lz_*^2}{z_0^3}(\frac{2}{3}+\beta^2\gamma^2)
\end{gather}
 
The surface integral is just the integral of $\eta^1$ over the co-dimension two minimal surface ($\mathcal S$). Let us evaluate the integrals one at a time. The first term in the homogeneous part can be calculated easily and gives,
\begin{gather}
\int_{-L}^L \int_0^\pi\frac{C_1\cos(\theta)}{\sqrt{\sin(\theta)}}\frac{1}{2z_* \sin^{\frac{3}{2}}(\theta)}d\theta~d\lambda=-\frac{LC_1}{z_*}\left[\frac{1}{\sin\theta}\right]^{\theta\rightarrow\pi}_{\theta\rightarrow 0}
\end{gather}
The second term in the homogeneous part has to be evaluated by an integration by parts. An integration by parts yields the following expression,
\begin{gather}
\int_{-L}^L \int_0^\pi C_2\sin{\theta}~~_2F_1\bigg(\frac{1}{4}, 1; \frac{1}{2}, \cos^2\theta\bigg) \frac{1}{2z_* \sin^{\frac{3}{2}}(\theta)} d\theta=\frac{LC_2}{z_*}\left[-\sin^{\frac{3}{2}}\theta~~_2F_1\bigg(\frac{1}{4}, 1; \frac{1}{2}, \cos^2\theta\bigg)\frac{\cos\theta}{\sin\theta}\right]^{\theta\rightarrow\pi}_{\theta\rightarrow 0}\notag\\
+\frac{LC_2}{z_*}\int_0^\pi\frac{d}{d\theta}\left(\sin^{\frac{3}{2}}\theta~~_2F_1\bigg(\frac{1}{4}, 1; \frac{1}{2}, \cos^2\theta\bigg)\right)\frac{\cos\theta}{\sin\theta}d\theta
\end{gather}
There are divergences in the integrals of both the homogeneous terms. We will therefore not evaluate it right away, but show how the divergences cancel with the terms arising from the particular solution. To evaluate the pieces in the particular solution, let us define the function,
\begin{gather}
f(\theta)=\left[\frac{1}{4}\left(3D+\frac{3C}{2}\right)z_*^3\left(\sin{\theta}\right)^{\frac{1}{2}}-\frac{7}{8}Dz_*^3\left(\sin{\theta}\right)^{\frac{5}{2}}\right](\sin{\theta})^{\frac{3}{2}}~~_2F_1\bigg(\frac{1}{4}, 1; \frac{1}{2}, \cos^2\theta\bigg)
\end{gather}
The first term in the particular solutions can then be written as, 
\begin{gather}
-\int_0^\pi\frac{2L}{2z_*\sin{\theta}^{\frac{3}{2}}}\frac{\cos(\theta)}{\sqrt{\sin(\theta)}}\int_0^\theta f(\theta')d\theta'=\frac{L}{z_*}\left[\frac{\int_0^\theta f(\theta')d\theta'}{\sin\theta}\right]_{\theta\rightarrow 0}^{\theta\rightarrow\pi}-\frac{L}{z_*}\int_0^\pi \frac{f(\theta)}{\sin\theta}d\theta
\end{gather}
By virtue of the boundary conditions and consequently the values $C_1$ and $C_2$, the terms diverging as $\frac{1}{\sin\theta}$ cancel and we are left with,
\begin{gather}
\frac{LC_2}{z_*}\int_0^\pi\frac{d}{d\theta}\left[\sin^{\frac{3}{2}}\theta~~_2F_1\bigg(\frac{1}{4}, 1; \frac{1}{2}, \cos^2\theta\bigg)\right]\frac{\cos\theta}{\sin\theta}d\theta-\frac{L}{z_*}\int_0^\pi \frac{f(\theta)}{\sin\theta}d\theta
\end{gather}

The first of the above two integrals then evaluates to $-\frac{\Gamma \left(\frac{1}{4}\right)^3 Lz_*^2\left(2C+D\right)}{32~ \Gamma \left(\frac{3}{4}\right)\sqrt{2}}$ \footnote{Consider the expression $\sin^{\frac{3}{2}}\theta~~_2F_1(\frac{1}{4}, 1; \frac{1}{2}, \cos^2\theta)$ and let $\cos^2\theta=z$. Then the expression becomes $(1-z)^{\frac{3}{4}}~_2F_1(\frac{1}{4}, 1; \frac{1}{2}, z)$. Using Euler's identity, $_2F_1(a, b; c, z)=(1-z)^{c-a-b}~_2F_1(c-a,c-b;c,z)$ we have $_2F_1(\frac{1}{4}, -\frac{1}{2}; \frac{1}{2}, z)=(1-z)^{\frac{3}{4}}~_2F_1(\frac{1}{4}, 1; \frac{1}{2}, z)$
Therefore
$
\frac{d}{d\theta}\left(\sin^{\frac{3}{2}}\theta~_2F_1(\frac{1}{4}, 1; \frac{1}{2}, \cos^2\theta)\right)= -\tan \theta\left\{\, _2F_1\left(-\frac{1}{2},\frac{1}{4};\frac{1}{2};\cos ^2\theta\right)-(1-\cos ^2\theta)^{-\frac{1}{4}}\right\}$.}
and the last $f(\theta)$ integral can be calculated easily and is given as $-\frac{Lz_*^2}{10} (15 C+16 D)$.
The integral of the second particular solution of $\mathcal S$ gives $\frac{2Lz_*^2}{10}  (5 C+8 D)$.
The total change in complexity therefore turns out to be,
\begin{eqnarray}
\Delta C
={1\over 8\pi G}\frac{Lz_*^2}{2z_0^3} \left(1+\beta ^2 \gamma ^2-\frac{\pi \Gamma \left(\frac{5}{4}\right)^2 \left(1+2 \beta ^2 \gamma ^2\right)}{\Gamma \left(\frac{3}{4}\right)^2}\right)
\end{eqnarray}
 This result is consistent with the $4$- dimensional result for the `strip perpendicular to boost' case of \cite{Karar:2019bwy}. This is because of the different notions of parallel and perpendicular used there.

\subsubsection*{\underline{ Case B: Strip along `x' boost along `y'}}\label{SAXBAY}
To handle this case we will just change the direction of the boost rather than changing the orientation of the strip. Therefore, the embeddings, eq. (\ref{embstrip}) remains intach while $A$ gets replaced by $\tilde A$ and so on. Hence the solution for $\eta^1$, in this case, is same as that for eq. (\ref{SAXBAXSOL}) with $C$ replaced by $\tilde C$ and $D$ by $\tilde D$ (appendix (\ref{FG4})). The final expression for the change in complexity is obtained just by making this replacement and is given by.

\begin{equation}
\Delta C={1\over 8\pi G}\frac{Lz_*^2}{2z_0^3}\bigg(1+2\beta ^2 \gamma ^2-\frac{\pi \Gamma \left(\frac{5}{4}\right)^2 \left(1+\beta ^2 \gamma ^2\right)}{~\Gamma \left(\frac{3}{4}\right)^2}\bigg)
\end{equation}

This result is consistent with the $4$-dimensional result obtained in \cite{Karar:2019bwy}, but now for the `strip parallel to boost' case.

\section{Other Applications}
\subsection{Holographic entanglement entropy}
The general expressions found in section \ref{sop} have many applications when it comes to finding identities for the change in `entanglement entropy'. Consider the case of the change in entanglement entropy when the boundary subregion is deformed. It may be a spatial deformation of the subregion or an evolution in time.Some noted works in this direction can be found in \cite{Faulkner:2015csl,Carmi:2015dla,Cavini:2019wyb}. The only contribution to the change then comes from the integration over the boundary of the subsytem itself \footnote{Note that $\mathcal S$ plays the role of $\mathscr S$ and $\partial A$ plays the role of $\partial \mathscr S$ here. }.
\begin{eqnarray}
\delta S=-\frac{1}{4G}\int_{\partial A}g(\xi,\hat r') \sqrt{\zeta}~d^{d-3}\eta,
\end{eqnarray} 
where $\eta$ are coordinates on the boundary of the subsytem $\partial A$ and $\hat r'$ is the co-normal of $\partial A$ in $\mathcal S$. Note the integration over $A$ was absent in the previous expressions because the subsytem was kept fixed. Further, the solution of the homogeneous version of eq.(\ref{IJE}) is necessarily not a trivial solution because of non-trivial boundary conditions. We will demonstrate this for the $AdS_3$ case. Consider the change of enatnglement entropy for a particular deformation of the boundary subregion. Recall, that the solution of the homogeneous Jacobi equation in this case is given by \cite{Ghosh:2017ygi},
\begin{eqnarray}
N^{0}=C_1e^\tau+C_2e^{-\tau},~~N^1=B_1e^\tau+B_2e^{-\tau},
\end{eqnarray}
where $\tau$ is the parameter along the space-like geodesic. The tangent component $N^2$ in this case will be non-zero. However it is completely characterised by it's boundary value. While the normal components $N^{0}$ and $N^{1}$ are to be obtained by solving the homogeneous Jacobi equation with boundary conditions determined by nature of deformation of subregion $A$. The boundary $\partial A$ has two disjoint points. Let us denote them as $\partial A_1$ and $\partial A_2$. It follows that the deformation vector $X$ (say) is given by (assuming there is no deformation into the bulk $X^z=0$). The deformation vector field ($X=A(x,t)\partial_t+B(x,t)\partial_x$) is then related to the solutions of the homogeneous Jacobi equation through the following relations. For the ingoing branch,
\begin{gather}
g(X,e_0)=\frac{A}{z_*\sech{\tau}}\nn
~\implies~\lim_{\tau\rightarrow-\infty}z_*\sech(\tau)N^0=A\bigg(-\frac{l}{2},t\bigg)=2z_*c_2,~~\lim_{\tau\rightarrow\infty}\sech(\tau)N^0=A\bigg(\frac{l}{2},t\bigg)=2z_*c_1\nn
g(X,e_2)=\frac{\tanh{\tau}}{z_*\sech{\tau}}~B~\nn
\implies~\lim_{\tau\rightarrow-\infty}\frac{z_*\sech{\tau}}{\tanh{\tau}}N^2=B\bigg(-\frac{l}{2},t\bigg)=-2z_*b_2,~~\lim_{\tau\rightarrow\infty}\frac{z_*\sech{\tau}}{\tanh{\tau}}N^2=B\bigg(\frac{l}{2},t\bigg)=2z_*b_1\nn
g(X,e_1)=\frac{B}{z_*}~\implies~N^1=\frac{B}{z_*}
\end{gather}
 For the $2+1$ dimensionl case the expression for the change of entanglement entropy becomes,
\begin{eqnarray}
\delta S&=&\frac{1}{4G}g(T,N)\bigg|_{\partial A}\nn
\delta^2S&=&\frac{1}{4G}\left(g(\nabla_NN,T)+\frac{1}{2}\nabla_Tg(N^\perp,N^\perp)\right)\bigg|_{\partial A}
\end{eqnarray}

The constants $B,C$ are therefore fixed in terms of the components of the vector deforming the boundary subregion. For a general asymptotically $AdS$ consider it's conformal completion $\tilde g=\Omega^2~ g$. The asymptotic boundary is the given by $\Omega=0$. The normal is then $n_a=\nabla_a\Omega$. The deformation vector field can then be considered to purely tangential to the $\Omega=constant$ hypersurfaces. On such a hypersurface one has the following decomposition,
\begin{eqnarray}
\nabla_NN=\mathscr D_NN+\mathscr{K}(N,N)+\frac{1}{2\Omega^2}g(N,N)\nabla\Omega^2, 
\end{eqnarray}
where $\mathscr D$ and $\mathcal {K}$ are the intrinsic covariant derivative and extrinsic curvature of a constant $\Omega$ hypersurface. The second variation the reduces to,

\begin{eqnarray}
\delta^2S&=&\frac{1}{4G}\left(g(\mathscr D_NN+\mathscr{K}(N,N),T)+\frac{1}{2\Omega^2}\nabla_T\left(\Omega^2g(N^\perp,N^\perp)\right)+\frac{g(N^T,N^T)}{2\Omega^2}\nabla_T\Omega^2\right)\bigg|_{\partial A}
\end{eqnarray}
The last two terms has to be understood as limits. For the Poincare patch of pure $AdS$, $\Omega=z$. The overall variation turns out to be,
\begin{eqnarray}
\Delta S=\frac{1}{4G}\bigg(2(b_1-b_2)^2-4(c_1-c_2)^2-2\left(\frac{B^2}{z_*^2}\right)\bigg|_{\partial A}\bigg)
\end{eqnarray}
For a time-like deformation, more specifcally a constant constant time translation of the subregion could be initiated by $X=A\partial_t$. In such a case $b_1=b_2=0$ and $ c_1=c_2$, which leads to the result $\Delta S=0$. For a increase in the subsytem size by $\frac{\delta l}{2}$ on each side, one has, $b_1=\frac{\delta l}{4 z_*},~b_2=-\frac{\delta l}{4 z_*}$. The change in entropy is then given by,
\begin{eqnarray}
\Delta S=\frac{\delta l^2}{4G}
\end{eqnarray}
While the approach still holds for higher dimensions, unfortunately, it will give divergent results. This because in higher dimensions the cut off dependent term in entanglement entropy itself depends on the size of the subsytem. Hence any infinitesimall change in the size will give divergent changes in entanglement entropy. This can however be cured if the integrals in the above approach is done on a bulk cut-off surface $z=\epsilon$ (say).
\section{Conclusion}
From the information theoretic or computational perspective complexity is an important quantity. It needs to be determined what are the resources needed to perform a particular computational task. In a quest for quantum computation, it has emerged as highly relevant, when it comes to determining the states that allow for fast computation. Holographic principles can be effectively used to calculate the complexity of states that have a dual. This has reduced the daunting task of calculating complexity in field theories to the calculation of certain geometric quantities. In spite of this simplification the determination of subregion complexity geometrically amounts to solving a second order elliptic differential equation. Though, this is possible for a few bulk geometries which are highly symmetric, for more general space-times solving such an equation is difficult. The general consensus is then to do it perturbatively around the solutions known for pure $AdS$.

One of the approaches is to identify a parameter of the bulk space-time and perform an asymptotic expansion of the volume integral in terms of the parameter. The asymptotic limit of the parameter gives the $AdS$ solution. But this is inherently problem specific and  the expansion parameter of-course depends on the bulk. Our approach on the other hand is a variational one that computes terms in order of the strength of a perturbation, which is conveniently taken to be in the Fefferman-Graham gauge. Our approach is in no way advantageous over the previous one in terms of algebraic simplification. However, the advantage is that it provides universal expressions applicable to any space-time that is a perturbation over $AdS$ or for that matter any back-ground. What it does is to cast the problem in a form that might give further insights.

The expression eq. (\ref{firstvariation}) for example refers to a general perturbation $P(\partial_a,\partial_b)$ and the restriction of the deviation vector $N$ to the boundary $\partial\mathscr S$. The restriction of the deviation vector $N$ to the boundary $\partial\mathscr S$, $\eta$ satisfies a linear elliptic equation. There is considerable amount of literature in the theory of differential equations of the elliptic type. We hope that various bounds on the solutions of equations of this type may yield interesting inequalities in the context of complexity and entanglement entropy.

It is important to note what the ingredients required to calculate the first order change are and how it compares to the known approaches. It is clear that one requires the pull back of the perturbation $P$ on $\Sigma$ and the deviation of $\mathcal S$ along $\Sigma$ or the initial constant $t$ slice of pure $AdS$. It might therefore seem that even by restricting oneself to a $\Sigma$ such calculations can be done. In other words deviations away from the initial constant $t$ slice are redundant and may just consider a co-dimension one Jacbi equation on $\Sigma$ . But this may not be completely correct. For example it is not known whether the component of the deviation vector along $\hat r$ satisfies a co-dimension one Jacobi equation on $\Sigma$ itself (and not on the full spacetime $\mathcal M$). If the answer is in the negative then the full co-dimension two Jacobi equation on $\mathcal S$ must be solved, as has been done in our case, in order to find the correct component along $\hat r$. Hence completely restricting oneself to $\Sigma$ may not be a good approximation after all.  

Finally, we believe that the approach taken by us might have important consequences in the context of bulk reconstruction. Specifically, in the absence of a perturbation, the solutions of the Jacobi equation tell us how the minimal surface deforms with the deformation of the boundary subregion. Hence the Jacobi equation tells us how the bulk is traced as one deforms the boundary subregion. This might be useful to find the flow of the metric induced on a given minimal surface, thus reconstructing a region of space-time.

\section{Acknowledgements}
Most of the integrals and solutions of the differential equations were done using Maple \cite{maple} and Mathematica \cite{mathematica}. Tensor calculations were done with the GRTensor package \cite{grtensor} on a Maple platform. AG would like to thank Sudipta Sarkar for helpful discussions. RM would like to thank Koushik Ray for helpful discussions. AG is supported by SERB, government of India through the NPDF grant (PDF/2017/000533).

\appendix

\section{Fefferman Graham Expansion}\label{FG}
\subsection{Asymptotic expansion of BTZ}\label{FG3}
The rotating BTZ metric is given as,
\begin{gather}
ds^2=-\frac{(r^2-r_+^2)(r^2-r_-^2)}{r^2}~dt^2
+\frac{r^2}{(r^2-r_+^2)(r^2-r_-^2)}~dr^2+r^2\left(dx-\frac{r_+r_-}{r^2}~dt\right)^2,\label{mBTZ}
\end{gather}
where $r_+,r_-$ are the radii of the outer and inner horizon respectively. An asymptotic expansion of the above metric can be realized by defining coordinate $\rho$ such that $\frac{d\rho}{\rho}=\frac{dr}{r\sqrt{f(r)}}$. In terms of $\rho$ this metric becomes,

\begin{gather}
ds^2=\frac{d\rho^2}{\rho^2}+\rho^2 \Biggl[(-dt^2+dx^2)+\frac{1}{\rho^2}\Biggl(\frac{(r_+^2 + r_-^2)}{2}dt^2 - 2 r_+ r_- dt dx+\frac{(r_+^2 + r_-^2)}{2}dx^2\Biggr)\nn
+\frac{1}{\rho^4}\Biggl(\frac{-(r_-^2-r_+^2)^2}{16}dt^2+\frac{(r_-^2-r_+^2)^2}{16}dx^2\Biggr)\Biggr]
\end{gather}
By writing the metric in coordinates $(\rho=\frac{1}{z})$, one can identify the perturbations as,
\[
\accentset{(1)}{h}_{\mu\nu}=
\begin{bmatrix}
\frac{(r_+^2 + r_-^2)}{2} & ~~0~~ &-r_+ r_- \\
0&0&0\\
-r_+ r_- & 0 & \frac{(r_+^2 + r_-^2)}{2}.
\end{bmatrix}
\]
The components of $C^{\mu}~_{\nu\rho}$ can be obtained with the expressions for $\accentset{(1)}{h}_{\mu\nu}$. Denoting $a=\frac{(r_+^2 + r_-^2)}{2}$ and $b=-r_+ r_-$ one has,
\begin{gather}
\accentset{(1)}{C} ^{t }~_{z ~t }=-z ~a,~\accentset{(1)}{C}^{x }~_{z ~t }=z ~b,~\accentset{(1)}{C} ^{t }~_{t ~z }=-z ~a,~\accentset{(1)}{C} ^{x}~_{t ~z }=z ~b\nn
\accentset{(1)}{C} ^{t }~_{x ~z }=-z ~b,~\accentset{(1)}{C}^{x }~_{x ~z }=z ~a,~{C} ^{t }~_{z ~x }=-z ~b,~\accentset{(1)}{C} ^{x }~_{z ~x }=z ~a
\end{gather}
 \subsection{Asymptotic expansion of Boosted Black brane}\label{FG4}
 The boosted black brane metric in holographic coordinates is of the form 
\begin{gather}
 ds^2={R^2\over z^2}\left[-\mathcal {A}(z)dt^2+\mathcal{B}(z)dx^2+\mathcal{C}(z)dtdx+dx^2+{dz^2\over f(z)}\right],
\end{gather}
where,
\begin{gather}
 \mathcal{A}(z)=1-\gamma^2({z\over z_0})^3,~~\mathcal{B}(z)=1+\beta^2\gamma^2({z\over z_0})^3,\\
 \mathcal{C}(z)=2\beta\gamma^2({z\over z_0})^3,~~f(z)=1-({z\over z_0})^3
\end{gather}
 $z_0$ is the location of the horizon while $0\leq\beta\leq1$ is the boost parameter. As usual $\gamma$ is defined as ${1\over\sqrt{1-\beta^2}}$. The boost here is along $x$ direction. By writing the metric in asymptotic coordinates one can view this as a perturbation over $AdS_4$. This is achieved by defining a coordinate $\rho$ such that,
\begin{gather}
 {dz\over z\sqrt{f(z)}}={d\rho\over\rho}
\end{gather}
Integrating this and setting the integration constant to $({\rho_0}^3=4{z_0}^3)$ we get,
\begin{gather}
 {1\over z^2}={1\over\rho^2}(1+({\rho\over{\rho_0}})^3)^{4\over 3}={1\over\rho^2}g(\rho)^{4\over 3}
\end{gather}
Now we expand the metric coefficient upto second order in $({\rho\over{\rho_0}})^3$, Substituting this back in the metric we get
\begin{gather}
 ds^2
 ={R^2\over\rho^2}\Biggl[d\rho^2+\left(\eta_{\mu\nu}+\rho^3\gamma^{(3)}_{\mu\nu}+\rho^6\gamma^{(6)}_{\mu\nu}\right)dx^{\mu}dx^{\nu}\Biggr]
\end{gather}
Where
\begin{gather}
 \gamma^{(3)}_{\mu\nu}=\begin{bmatrix}
        -({1\over 3}-\gamma^2)({1\over z_0})^3 & \beta\gamma^2({1\over z_0})^3 & 0 \\
          \beta\gamma^2({1\over z_0})^3 & \left({1\over 3}+\beta^2\gamma^2\right)({1\over z_0})^3 & 0\\
          0 & 0 & {1\over 3}({1\over z_0})^3
        \end{bmatrix}\label{got}
\end{gather}
One can check that $Tr( \gamma^{(3)}_{\mu\nu})=0$ and
\begin{gather}
 \gamma^{(6)}_{\mu\nu}=\begin{bmatrix}
        -\left({2\over 9}+{8\over 3}\gamma^2\right){1\over16 {z_0}^6} & -{1\over 6}\beta\gamma^2({1\over z_0})^6 & 0 \\
         -{1\over 6}\beta\gamma^2({1\over z_0})^6  & \left({2\over 9}-{8\over 3}\beta^2\gamma^2\right){1\over 16 {z_0}^6} & 0\\
          0 & 0 & {2\over 9}{1\over 16{z_0}^6}
        \end{bmatrix}
\end{gather}
The perturbation $\accentset{(1)}{P}_{\mu\nu}$  and $\accentset{(2)}{P}_{\mu\nu}$ can be read off as, $\accentset{(1)}{P}_{\mu\nu}=\gamma^{(3)}_{\mu\nu}z$ and $\frac{1}{2}\accentset{(2)}{P}_{\mu\nu}=\gamma^{(6)}_{\mu\nu}z^4$ respectively. Also, the following notation will be used in the main text $C=\left({1\over 3}+\beta^2\gamma^2\right){1\over z_0^3},~D={1\over 3}{1\over z_0^3}$.

For boost along the $y$ axis, $\accentset{(1)}{P}(\partial_\mu,\partial_\nu)$ and $\accentset{(2)}{P}(\partial_\mu,\partial_\nu)$ is of the form,

{\[\displaystyle \accentset{(1)}{P} _{\mu\nu}=\left(\begin{array}{cccc}\tilde A ~z  & 0 & \tilde B ~z  & 0 \\0 & \tilde C ~z  & 0 & 0 \\\tilde B ~z  & 0 & \tilde D~z  & 0 \\0 & 0 & 0 & 0 \\\end{array}\right)~~~~\displaystyle \frac{1}{2}\accentset{(2)}{P} _{\mu\nu}=\left(\begin{array}{cccc}\tilde A' ~z^4  & 0 & \tilde B' ~z^4  & 0 \\0 & \tilde C' ~z^4  & 0 & 0 \\\tilde B' ~z^4  & 0 & \tilde D'~z^4  & 0 \\0 & 0 & 0 & 0 \\\end{array}\right)\]}\\
 
where $\tilde{C}={1\over 3}({1\over z_0})^3,~~\tilde{D}=\left({1\over 3}+\beta^2\gamma^2\right)({1\over z_0})^3,~~ 
 \tilde{C}^\prime={2\over 9}{1\over 16{z_0}^6},~~\tilde{D}^\prime=\left({2\over 9}-{8\over 3}\beta^2\gamma^2\right){1\over 16 {z_0}^6},~~B=\tilde{B}=\beta\gamma^2({1\over z_0})^3$. 

\section{Second order volume variation}\label{sva}
 In this section we will briefly go through the derivation of the second variation of the area functional including metric perturbation and deviation of the surface. Starting from the area functional and computing its second variation gives the following expression,
 \begin{gather}
\delta^{2}A=\underbrace{\int d^{n}\tau~(\delta\sqrt{h})h^{ab}\left(g(\nabla_{\partial_{a}}N,\partial_{b})+{1\over 2}\accentset{(1)}{P}(\partial_{a},\partial_{b})\right)}_{\text{I}}+\underbrace{\int d^{n}\tau~\sqrt{h}(\delta h^{ab})\left(g(\nabla_{\partial_{a}}N,\partial_{b})+{1\over 2}\accentset{(1)}{P}(\partial_{a},\partial_{b})\right)}_{\text{II}}\\\notag
+\underbrace{\int d^{n}\tau~\sqrt{h}h^{ab}\delta\left(g(\nabla_{\partial_{a}}N,\partial_{b})+{1\over 2}\accentset{(1)}{P}(\partial_{a},\partial_{b})\right)}_{\text{III}}.
 \end{gather}
The above expression now has to be restructured so as to extract the first order inhomogeneous Jacobi equations. In order to achieve this we will first separate this expression into terms with perturbations and the ones without. We will then separate the terms containing the normal component of the deviation vector from those containing the tangent component. At the very end we will isolate the boundary terms from the bulk ones. To carry out this formidable task we mark the individual terms as I, II and III. Term I in the above expression can be written as,
 \begin{gather}
 \text{I}=\underbrace{\int d^{n}\tau~\sqrt{h}h^{cd}h^{ab}\Biggl[g(\nabla_{\partial_{c}}N,\partial_{d})g(\nabla_{\partial_{a}}N,\partial_{b})\Biggr]}_{A_1}\nonumber\\+\underbrace{\int d^{n}\tau~\sqrt{h}h^{cd}h^{ab}\Biggl[P(\partial_{a},\partial_{b})g(\nabla_{\partial_{c}}N,\partial_{d})+{1\over 4}h^{ab}h^{cd}P(\partial_{c},\partial_{d})\accentset{(1)}{P}(\partial_{a},\partial_{b})\Biggr]}_{A_2}\label{one1},
 \end{gather}
 where we have used the first variation of $h_{ab}$. On using the expression $h^{ac}h^{bd} P(\partial_{a},\partial_{b})g(\nabla_{\partial_{c}}N,\partial_{d})=P((\nabla_{\partial_{c}}N)^T,\partial_{a})$, term II can be written as, 
\begin{gather}
\text{II}=-\underbrace{\int d^{n}\tau~\sqrt{h}\left[h^{ac}h^{bd}(g(\nabla_{\partial_{c}}N,\partial_{d})+g(\nabla_{\partial_{d}}N,\partial_{c}))g(\nabla_{\partial_{a}}N,\partial_{b})\right]}_{B_1}\nonumber\\-\underbrace{\int d^{n}\tau~\sqrt{h}\left[2h^{ab}\accentset{(1)}{P}((\nabla_{\partial_{a}}N)^T,\partial_{b})+{1\over 2}h^{ac}h^{bd}\accentset{(1)}{P}(\partial_{a},\partial_{b})\accentset{(1)}{P}(\partial_{c},\partial_{d})\right]}_{B_2}\label{two2}
\end{gather}
Similarly term-III can be written as,
\begin{gather}
\text{III}=\underbrace{\int d^{n}\tau~\sqrt{h}h^{ab}\left(g(\nabla_N\nabla_{\partial_{a}}N,\partial_{b})+g(\nabla_{\partial_{a}}N,\nabla_{\partial_{b}}N)\right)}_{C_1}\nonumber\\+\underbrace{\int d^{n}\tau~\sqrt{h}h^{ab}\left(2\accentset{(1)}{P}(\nabla_{\partial_{a}}N,\partial_{b})+2g(C(\partial_a,N),\partial_b)+{1\over 2}\accentset{(2)}{P}(\partial_{a},\partial_{b})\right)}_{C_2}\label{three3}
\end{gather}
 From expression of eq. \eqref{one1}, eq. \eqref{two2} and eq. \eqref{three3} we can see that we have been successful in separating terms with perturbation from those without. Now we will try to further simplify these. Let us consider the terms without perturbation first.

\subsection*{Terms without perturbations}
The terms without perturbation have been grouped as $A_1~,B_1$ and $C_1$. Their sum is therefore total contribution from terms not containing perturbation quantities.
\begin{gather}
 A_1+B_1+C_1= \underbrace{\int d^{n}\tau~\sqrt{h}h^{cd}h^{ab}g(\nabla_{\partial_{c}}N,\partial_{d})g(\nabla_{\partial_{a}}N,\partial_{b})}_{(1)}\nonumber\\
 +\underbrace{\int d^{n}\tau~\sqrt{h}h^{ab}g(\nabla_{\partial_{a}}N,\nabla_{\partial_{b}}N)-\int d^{n}\tau~\sqrt{h}\left[h^{ac}h^{bd}(g(\nabla_{\partial_{c}}N,\partial_{d})+g(\nabla_{\partial_{d}}N,\partial_{c}))g(\nabla_{\partial_{a}}N,\partial_{b})\right]}_{(2)}\nonumber\\+\underbrace{\int d^{n}\tau~\sqrt{h}h^{ab}g(\nabla_N\nabla_{\partial_{a}}N,\partial_{b})}_{(3)}
\end{gather}
For our convenience we have seperated the terms and identified them as $(1),(2)$ and $(3)$. Now let us first consider the term $(2)$
\begin{gather}
 h^{ab}g(\nabla_{\partial_{a}}N,\nabla_{\partial_{b}}N)-\left[h^{ac}h^{bd}(g(\nabla_{\partial_{c}}N,\partial_{d})+g(\nabla_{\partial_{d}}N,\partial_{c}))g(\nabla_{\partial_{a}}N,\partial_{b})\right]\nonumber\\
=h^{ab}g((\nabla_{\partial_{a}}N)^{\perp},(\nabla_{\partial_{b}}N)^{\perp})-h^{ac}h^{bd}g(D_{\partial_{d}}N^T,\partial_{c})g(D_{\partial_{a}}N^T,\partial_{b})+2h^{ab}g(D_{\partial_{a}}N^T,W_{N^{\perp}}(\partial_{b}))\nonumber\\
-h^{ab}g(W_{N^{\perp}}(\partial_a),W_{N^{\perp}}(\partial_b)),
\end{gather}
where we have used the decompositions
$ \nabla_{\partial_a}N^T=D_{\partial_a}N^T+K(\partial_a,N^T);~\nabla_{\partial_a}N^{\perp}=-W_{N^{\perp}}(\partial_a)+\nabla^{\perp}_{\partial_a}N^{\perp}$. As described before $K(\partial_a,\partial_b)$ is the extrinsic curvature and $W_{N^{\perp}}(\partial_a)$ is the shape operator. They are related by the Weingarten equation $g(W_{N^{\perp}}(\partial_a))=g(N^{\perp},K(\partial_a,\partial_b))$. The shape operator van also be shown to be self adjoint i.e $g(W_{N^{\perp}}(\partial_a),\partial_b)=g(W_{N^{\perp}}(\partial_b),\partial_a)$. Using this we can write the above expression as
\begin{gather}
(2)=h^{ab}\Biggl(g(K(\partial_{a},N^T),K(\partial_{b},N^T))+2g(K(\partial_{a},N^T),\nabla^{\perp}_{\partial_{b}}N^{\perp})+g(\nabla^{\perp}_{\partial_{a}}N^{\perp},\nabla^{\perp}_{\partial_{b}}N^{\perp})\nonumber\\+2g(D_{\partial_{a}}N^T,W_{N^{\perp}}(\partial_{b}))-g(W_{N^{\perp}}(\partial_a),W_{N^{\perp}}(\partial_b)\Biggr)-h^{ac}h^{bd}g(D_{\partial_{d}}N^T,\partial_{c})g(D_{\partial_{a}}N^T,\partial_{b})
\end{gather}
In the above expression we have been able to seperate tangent contributions from normal contributions. Next we consider the term $(3)$ in the expression for $\delta^{2}_NA$
\begin{gather}
 (3)= h^{ab}g(\nabla_N\nabla_{\partial_{a}}N,\partial_{b})
=-[Ric(N^T,N^T)+2Ric(N^T,N^{\perp})+Ric(N^{\perp},N^{\perp})]\nonumber\\+h^{ab}[g(D_{\partial_a}(\nabla_{N^T}N^T)^T,\partial_b)-2g(D_{\partial_a}W_{N^{\perp}}(N^T),\partial_b)
+g(D_{\partial_a}([N^{\perp},N^T])^T,\partial_b)+g(D_{\partial_a}(\nabla_{N^{\perp}}N^{\perp})^T,\partial_b)]
\end{gather}
A similar exercise for term $(1)$ yields,
\begin{gather}
 \int d^{n}\tau~\sqrt{h}h^{cd}h^{ab}g(\nabla_{\partial_{c}}N,\partial_{d})g(\nabla_{\partial_{a}}N,\partial_{b})
=\int d^{n}\tau~\sqrt{h}h^{ab}h^{cd}g(D_{\partial_{a}}N^T,\partial_{b})g(D_{\partial_{c}}N^T,\partial_{d})
 \end{gather}
We see that the expression for $(1)$ has contributions only from tangent terms. Adding $(1),~(2)$ and $(3)$ we get the full expression. Note that the equations of motion $H=0$ and the Gauss and Codazzi identities \eqref{Iden} has been used. The full expression $ \delta^{2}_NA$ coming from terms that do not contain perturbation contributions is,

\begin{gather}\label{defo}
 \delta^{2}_NA=\int d^{n}\tau~\sqrt{h}\nabla_{\partial_a}[h^{ab}g(N^{\perp},{\nabla^{\perp}}_{\partial_{b}}N^{\perp})]-\int d^{n}\tau~\sqrt{h}g(\mathcal{L}(N^{\perp}),N^{\perp})
 \\+\int d^{n}\tau~\sqrt{h}h^{cd}g(D_{\partial_d}[N^Th^{ab}g(D_{\partial_a}N^T,\partial_b)],\partial_c)+\int d^{n}\tau~\sqrt{h}h^{ab}g(D_{\partial_a}([N^{\perp},N^T])^T,\partial_b)\\+\int d^{n}\tau~\sqrt{h}h^{ab}g(D_{\partial_a}(\nabla_{N^{\perp}}N^{\perp})^T,\partial_b),
\end{gather}
where $\mathcal{L}(N^{\perp})$ is the Jacobi operator defined in section \ref{soperte}.
Thus we have achieved the decomposition of the terms without perturbation in terms of tangent and normal contributions. Next we consider the terms with perturbations.
\subsection{Terms containing perturbations}
The terms containing perturbations have been grouped together. Thus the contribution to the area variation from terms containing perturbations are,
\begin{gather}
\delta^{2}_PA
=\int d^{n}\tau~\sqrt{h}\left[2h^{ab}g(\accentset{(1)}{C}(\partial_a,N),\partial_b)+2h^{ab}\accentset{(1)}{P}((\nabla_{\partial_{a}}N)^{\perp},\partial_{b})+h^{cd}h^{ab}\accentset{(1)}{P}(\partial_{c},\partial_{d})g(\nabla_{\partial_{a}}N,\partial_{b})\right]\\+\int d^{n}\tau~\sqrt{h}\left[{h^{ab}\over 2}\accentset{(2)}{P}(\partial_{a},\partial_{b})-{1\over 2}h^{ac}h^{bd}\accentset{(1)}{P}(\partial_{a},\partial_{b})\accentset{(1)}{P}(\partial_{c},\partial_{d})+{1\over 4}h^{ab}h^{cd}\accentset{(1)}{P}(\partial_{c},\partial_{d})\accentset{(1)}{P}(\partial_{a},\partial_{b})\right]
\end{gather}
Now our objective is to decompose the above expression in terms of tangent and normal contributions.
By using the identities in appendix \ref{Iden} we can write the above expression as
\begin{gather}
 \delta^{2}_PA
 =\int d^{n}\tau~\sqrt{h}h^{ab}\nabla_{\partial_a}[h^{cd}\accentset{(1)}{P}(\partial_c,\partial_d)g(N,\partial_b)]\nonumber\\
 +\int d^{n}\tau~\sqrt{h}\Biggl[2h^{ab}g(\accentset{(1)}{C}(\partial_a,N^{\perp}),\partial_b)+2h^{ab}\accentset{(1)}{P}((\nabla_{\partial_{a}}N^{\perp})^{\perp},\partial_{b})-h^{ab}h^{cd}\accentset{(1)}{P}(\partial_{c},\partial_{d})g(N,\nabla_{\partial_{a}}\partial_{b})\Biggr]\nonumber\\
 +\int d^{n}\tau~\sqrt{h}\Bigl[{h^{ab}\over 2}\accentset{(2)}{P}(\partial_{a},\partial_{b}) -{1\over 2}h^{ac}h^{bd}\accentset{(1)}{P}(\partial_{a},\partial_{b})\accentset{(1)}{P}(\partial_{c},\partial_{d})+{1\over 4}h^{ab}h^{cd}\accentset{(1)}{P}(\partial_{c},\partial_{d})\accentset{(1)}{P}(\partial_{a},\partial_{b})\Bigr]
\end{gather}
Now our task is to decompose the above expression further in terms of bulk and boundary contributions
\begin{gather}
 (\nabla_{\partial_a}\accentset{(1)}{P})(N^{\perp},\partial_b)=g(\accentset{(1)}{C}(\partial_a,N^{\perp}),\partial_b)+g(\accentset{(1)}{C}(\partial_a,\partial_b),N^{\perp})
\end{gather}
Substituting this we get the term containing $\accentset{(1)}{C}$ as
\begin{gather}
 h^{ab}(\nabla_{\partial_a}\accentset{(1)}{P})(N^{\perp},\partial_b)+h^{ab}\accentset{(1)}{P}((\nabla_{\partial_a}N^{\perp})^{\perp},\partial_b)=h^{ab}\nabla_{\partial_a}[\accentset{(1)}{P}(N^{\perp},\partial_b)]+ h^{ab}h^{cd}\accentset{(1)}{P}(\partial_b,\partial_d)g(N^{\perp},K(\partial_a,\partial_c))\nonumber\\-h^{ab}\accentset{(1)}{P}(N^{\perp},(\nabla_{\partial_a}\partial_b)^{T})
\end{gather}
Now we should note that 
\begin{gather}
 \nabla_{\partial_a}[\sqrt{h}h^{ab}\accentset{(1)}{P}(N^{\perp},\partial_a)]
=\sqrt{h}h^{ab}\nabla_{\partial_a}[\accentset{(1)}{P}(N^{\perp},\partial_b)]-\sqrt{h}h^{ab}\accentset{(1)}{P}(N^{\perp},(\nabla_{\partial_a}\partial_b)^T)
\end{gather}
Which is same as the term $\accentset{(2)}{P}$. Similarly,
\begin{gather}
 \nabla_{\partial_a}[\sqrt{h}h^{ab}h^{cd}\accentset{(1)}{P}(\partial_c,\partial_d)g(N,\partial_b)]
 =\sqrt{h}h^{ab}\nabla_{\partial_a}\Bigl[h^{cd}\accentset{(1)}{P}(\partial_c,\partial_d)g(N,\partial_b)\Bigr]-\sqrt{h}h^{ab}h^{cd}\accentset{(1)}{P}(\partial_c,\partial_d)g(N,\nabla_{\partial_a}\partial_b)
\end{gather}
Which is same as the term $\accentset{(1)}{P}$. Where we have used the fact that $([\partial_a,\partial_b]=0)$. Substituting this back in the expression for $\delta^{2}_PA$ we get,
\begin{gather}
 \delta^{2}_PA=\int d^{n}\tau~\partial_a[\sqrt{h}h^{ab}h^{cd}\accentset{(1)}{P}(\partial_c,\partial_d)g(N,\partial_b)]+2\int d^{n}\tau~\partial_a[\sqrt{h}h^{ab}\accentset{(1)}{P}(N^{\perp},\partial_b)]\notag\\
 +2\int d^{n}\tau~\sqrt{h} \Bigl(h^{ab}h^{cd}\accentset{(1)}{P}(\partial_b,\partial_d)g(N^{\perp},K(\partial_a,\partial_c))-h^{ab}g(\accentset{(1)}{C}(\partial_a,\partial_b),N^{\perp})\Bigr)\notag\\
 +\int d^{n}\tau~\sqrt{h}\Bigl[{h^{ab}\over 2}\accentset{(2)}{P}(\partial_{a},\partial_{b})-{1\over 2}h^{ac}h^{bd}\accentset{(1)}{P}(\partial_{a},\partial_{b})\accentset{(1)}{P}(\partial_{c},\partial_{d})+{1\over 4}h^{ab}h^{cd}\accentset{(1)}{P}(\partial_{c},\partial_{d})\accentset{(1)}{P}(\partial_{a},\partial_{b})\Bigr]\label{wdefo}
\end{gather}
Thus we have been able to decompose the expression into terms containing normal and tangent contribution and further we have separated the terms contributing to the boundary and bulk. Next we will add the contribution from the terms with and without perturbation to obtain the total variation.
\subsection{Total variation}
 The expression for the total variation upto second order is obtained by adding eq. \eqref{defo} and eq.\eqref{wdefo}. It ontains terms which involve $\nabla_NN$. We would however like to write it in terms of $\nabla_{N^\perp}N^\perp$. This can be done with the following steps,
\beq
\nabla_N N=\nabla_{N^{\perp}} N^\perp+\nabla_{N^{\perp}} N^T+\nabla_{N^T} N^{\perp}+\nabla_{N^T} N^T
\eeq
Note that the third surface term can be written as,
\beq
h^{ab}g(D_{\partial_a}N^T,\partial_b)g(N^T,\hat{r})&=&\gamma^{AB}g(D_{\partial_A}N^T,\partial_B)g(N^T,\hat{r})+g(\nabla_{N^T}N^T,\hat{r})\\\nonumber
&=&\gamma^{AB}g(N^T,K^{(\mathcal S\hookrightarrow\Sigma)}(\partial_A,\partial_B))g(N^T,\hat{r})+g(\nabla_{N^T}N^T,\hat{r})
\eeq
Denote by $K^{(\mathcal S\hookrightarrow\mathcal M)}(\partial_A,\partial_B)$ as the extrinsic curvature of $\mathcal S$ in $\mathcal M$. Similarly denote the other extrinsic curvature. Then from Gauss's decomposition it follows that,  
\beq
K^{(\mathcal S\hookrightarrow\mathcal M)}(\partial_A,\partial_B)=K^{(\mathcal S\hookrightarrow\Sigma)}(\partial_A,\partial_B)+K^{(\Sigma\hookrightarrow\mathcal M)}(\partial_A,\partial_B)
\eeq
Since $\mathcal S$ is minimal in $\mathcal M$ we have,
\beq
\gamma^{AB}K^{(\mathcal S\hookrightarrow\mathcal M)}(\partial_A,\partial_B)=\gamma^{AB}K^{(\mathcal S\hookrightarrow\Sigma)}(\partial_A,\partial_B)=\gamma^{AB}K^{(\Sigma\hookrightarrow\mathcal M)}(\partial_A,\partial_B)=0
\eeq
from which it follows that $K^{(\Sigma\hookrightarrow\mathcal M)}(\hat r,\hat r)=0$, since $h^{ab}K^{(\Sigma\hookrightarrow\mathcal M)}(\partial_a,\partial_b)=0$. The following relations can be obtained by using this constraint.
\beq
g([N^{\perp},N^T],\hat r)=g(\nabla_{N^\perp}N^T,\hat r)+g(N^\perp,K^{(\Sigma\hookrightarrow\mathcal M)}(\hat r,\hat r))g(N^T,\hat r)
\eeq
\beq
g(\nabla_{N^\perp}N^\perp,\hat r)=g(\nabla_N N-\nabla_{N^\perp}N^T-g(\nabla_{N^T}N^T,\hat r)+g(N^\perp,K^{(\Sigma\hookrightarrow\mathcal M)}(\hat r,\hat r))g(N^T,\hat r)
\eeq
Finally putting everything back into the total variation one obtains the following expression. This expression is more useful because it has been expressed in terms of quantities which are directly known from boundary conditions or as solutions of the IJEs.
\begin{gather}
 \delta^{2}A=\int d^{n}\tau~\sqrt{h} \Bigl(h^{ab}h^{cd}\accentset{(1)}{P}(\partial_b,\partial_d)g(N^{\perp},K(\partial_a,\partial_c))-h^{ab}g(\accentset{(1)}{C}(\partial_a,\partial_b),N^{\perp})\Bigr)\nonumber\\
 +\int d^{n}\tau~\sqrt{h}\Bigl[{h^{ab}\over 2}\accentset{(2)}{P}(\partial_{a},\partial_{b})-{1\over 2}h^{ac}h^{bd}\accentset{(1)}{P}(\partial_{a},\partial_{b})\accentset{(1)}{P}(\partial_{c},\partial_{d})+{1\over 4}h^{ab}h^{cd}\accentset{(1)}{P}(\partial_{c},\partial_{d})\accentset{(1)}{P}(\partial_{a},\partial_{b})\Bigr]\nonumber\\
 +\int d^{n}\tau~\partial_a[\sqrt{h}h^{ab}h^{cd}\accentset{(1)}{P}(\partial_c,\partial_d)g(N,\partial_b)]+2\int d^{n}\tau~\partial_a[\sqrt{h}h^{ab}\accentset{(1)}{P}(N^{\perp},\partial_b)]\nonumber\\
 +\int d^{n}\tau~\sqrt{h}h^{ab}g(D_{\partial_a}(\nabla_{N}N)^T,\partial_b)+\int d^{n}\tau~\nabla_{\partial_a}[\sqrt{h}h^{ab}g(N^{\perp},{\nabla^{\perp}}_{\partial_{b}}N^{\perp})]
\end{gather}
\section{Second order Jacobi equation}
In this appendix we will try to give a brief derivation of the second order Jacobi equation. Since the first order Jacobi equation is obtained by taking first variation of the mean curvature $(H)$ and imposing $H=0$. Similarly the second order Jacobi equation is obtained by taking a second variation of the mean curvature. Now the first variation of the mean curvature is given by
\begin{equation}\label{appon}
 (\nabla_N+\delta_P) H=h^{ab}{L}_{ab}^\perp-g(H,(\nabla_{\partial_c}N)^\perp )h^{cd}\partial_d+\nabla_{N^T}H
\end{equation}
Where
\begin{equation}
 L_{ab}=B_{ab}-D_{ab}+S^{(N)}_{ab}=\nabla_{\partial_a}\nabla_{\partial_b}N+R(N,\partial_a)\partial_b-\nabla_{(\nabla_{\partial_a}\partial_b)^T} N+2 h^{cd}K(\partial_a,\partial_c)g(N,K(\partial_b,\partial_d))
\end{equation}
Note that in general $[N^\perp, \partial_a]\neq 0$. However in this appendix for simplicity we are only considering variation where $[N^{\perp},\partial_a]=0$ and assume that $N^{\perp}$  is perpendicular to the surface $\mathcal S$. In appendix \eqref{tanvar} we will show that the commutator terms actually cancel among each other. One can easily check from eq. \eqref{appon} that by using ($H=0$)condition we get back the first order Jacobi equation. Now we need to take a first variation of eq. \eqref{appon} to obtain the second order Jacobi equation. 
\begin{equation}\label{devi1}
 (\nabla_N+\delta_P)\left(h^{ab}{L}_{ab}^\perp-g(H,(\nabla_{\partial_c}N)^\perp )h^{cd}\partial_d+\nabla_{N^T}H\right)=(\nabla_{N^\perp}+\delta_P)\left(h^{ab}{L}_{ab}^\perp\right)
\end{equation}
Where we have used the on shell condition ($H=0$). Thus our first task is to take the covariant derivative of the quantities $B^{\perp}_{ab}, D^{\perp}_{ab}$ and $S^{(N)}_{ab}$ respectively.

\subsubsection{Second order Jacobi equation without perturbation} We will drop the superscript from $N^\perp$, to avoid clutter. For convenience let  $J:=\nabla_N N$. First we will start with $B_{ab}=\nabla_{\partial_a}\nabla_{\partial_b}N+R(N,\partial_a)\partial_b$. Thus
\begin{gather}
h^{ab}\nabla_N B_{ab}^\perp=h^{ab}\nabla_N(B_{ab}-B_{ab}^T)=h^{ab}\nabla_N(B_{ab}-g(B_{ab},\partial_c)h^{cd}\partial_d)\notag\\
=h^{ab}M_{ab}^\perp-h^{ab}g(({B}_{ab})^\perp,(\nabla_{\partial_c}N)^\perp)h^{cd}\partial_d-(\nabla_G N)^\perp.\label{bicon}
\end{gather}
Since we can only work with the tangent vectors, the normal part is obtained by subtracting the tangent part from the whole. The rest of the expression above follows from straight forward computation of the covariant derivatives. In the above expression the quantities $M_{ab}$ and $G$ are given below
\begin{gather}
 M_{ab}=\biggl[4 R(N,\partial_a)\nabla_{\partial_b}N+(\nabla_{\partial_a }R)(N,\partial_b)N+(\nabla_N R)(N,\partial_a)\partial_b+R(N,\nabla_{\partial_a}\partial_b)N\biggr.\notag\\ \biggl.+\nabla_{\partial_a}\nabla_{\partial_b}J+R(J,\partial_a)\partial_b\biggr],~
 G=h^{ab}B_{ab}^T=h^{ab}g(B_{ab},\partial_c)h^{ce}\partial_e.\label{bi2con}
\end{gather}
In the same manner we can take covariant derivative of $D_{ab}=\nabla_{(\nabla_{\partial_a}\partial_b)^T} N$
\begin{equation}\label{dicon}
 h^{ab}\nabla_N {D}_{ab}^\perp=h^{ab}\nabla_N(D_{ab}-{D^T}_{ab})=h^{ab}Q_{ab}^\perp+h^{ab}\big(\nabla_{L^T_{ab}}N\big)^\perp-h^{ab}g(D_{ab}^\perp,(\nabla_{\partial_c}N)^\perp)h^{cd}\partial_d-(\nabla_F N)^\perp
\end{equation}
As before this expression is obtained by subtracting the derivative of the tangent part from the derivative of the total. In order to achieve this form the following identity has been used $[N,(\nabla_{\partial_a}\partial_b)^T]=L^{T}_{ab}$. In the above expression the form of $Q_{ab}$ and $F$ is given below
\begin{equation}\label{di2con}
 Q_{ab}=R(N,(\nabla_{\partial_a}\partial_b)^T)N+\nabla_{(\nabla_{\partial_a}\partial_b)^T}N,~~~
 F=h^{ab}D^{T}_{ab}=h^{ab}g(D_{ab},\partial_c)h^{ce}\partial_e
\end{equation}
The last term is the Simon's term $S^{(N)}_{ab}=-h^{cd}K(\partial_a,\partial_c)\bigg[g(\nabla_{\partial_b}N,\partial_d)+g(\partial_b,\nabla_{\partial_d}N)\bigg]$ this is a normal vector hence its covariant derivative is obtained directly and the following lengthy expression is obtained
\begin{align}\label{sicon}
&h^{ab}\nabla_N S^{(N)}_{ab}=2h^{ab}h^{cd}g(N,K(\partial_b,\partial_d)) L_{ac}^\perp-h^{ab}g\bigg(S^{(N)}_{ab},(\nabla_{\partial_e}N)^\perp\bigg)h^{ef}\partial_f+h^{ab}S^{(J^{\perp})}_{ab}\notag\\&-2h^{ab}h^{cd}K(\partial_a,\partial_c)\bigg(g(R(N,\partial_b)N,\partial_d)+g(\nabla_{\partial_b}J^T,\partial_d)+g(\nabla_b N,\nabla_d N)\bigg)
\end{align}
Following results from eq.\eqref{bicon}, \eqref{dicon} and eq. \eqref{sicon} above we see that
\begin{align}\label{jacob}
&h^{ab}(\nabla_NL^{\perp}_{ab})=h^{ab}\left(\nabla_N B_{ab}^\perp-\nabla_N {D}_{ab}^\perp+\nabla_N S^{N}_{ab}\right)
=h^{ab}\left(M_{ab}^\perp-Q_{ab}^\perp+S^{(J^{\perp})}_{ab}\right)-2h^{ab}\big(\nabla_{L^T_{ab}}N\big)^\perp\notag\\&+2h^{ab}h^{cd}g(N,K(\partial_b,\partial_d)) L_{ac}^\perp-2h^{ab}h^{cd}K(\partial_a,\partial_c)\bigg(g(R(N,\partial_b)N,\partial_d)+g(\nabla_{\partial_b}J^T,\partial_d)+g(\nabla_b N,\nabla_d N)\bigg)
\end{align}
Where we have used the first order Jacobi equation$(h^{ab}L^{\perp}_{ab}=0)$ in obtaining the above result. Now one can further simplify this by using eq\eqref{bi2con},\eqref{di2con} and the onshell condition $(H=0)$ to obtain the following expression
\begin{gather}
 h^{ab}(\nabla_NL^{\perp}_{ab})=h^{ab}\left((\nabla_{\partial_a}\nabla_{\partial_b}J)^{\perp}-(\nabla_{(\nabla_{\partial_a}\partial_b)^T} J)^{\perp}+(R(J,\partial_a)\partial_b)^{\perp}+{S^{J}}_{ab}\right)+h^{ab}\biggl(4 R(N,\partial_a)\nabla_{\partial_b}N\biggr.\notag\\ \biggl.+(\nabla_{\partial_a }R)(N,\partial_b)N+(\nabla_N R)(N,\partial_a)\partial_b\vphantom{4 R(N,\partial_a)\nabla_{\partial_b}N}\biggr)^{\perp}-2h^{ab}\big(\nabla_{L^T_{ab}}N\big)^\perp+2h^{ab}h^{cd}g(N,K(\partial_b,\partial_d)) L_{ac}^\perp\notag\\-2h^{ab}h^{cd}K(\partial_a,\partial_c)\biggl(g(R(N,\partial_b)N,\partial_d)+g(\nabla_{\partial_b}J^T,\partial_d)+g(\nabla_b N,\nabla_d N)\biggr)\label{theone}
\end{gather}
Finally one can check that 
\begin{gather}
 \nabla_N\left(h^{ab}{L}_{ab}^\perp\right)=h^{ab}(\nabla_NL^{\perp}_{ab})+(\nabla_N h^{ab})L^{\perp}_{ab}=h^{ab}L^{(J)}_{ab}+h^{ab}\biggl(4 R(N,\partial_a)\nabla_{\partial_b}N+(\nabla_{\partial_a }R)(N,\partial_b)N\biggr.\notag\\ \biggl.+(\nabla_N R)(N,\partial_a)\partial_b\vphantom{4 R(N,\partial_a)\nabla_{\partial_b}N}\biggr)^{\perp}-2h^{ab}\big(\nabla_{L^T_{ab}}N\big)^\perp+4h^{ab}h^{cd}g(N,K(\partial_b,\partial_d)) L_{ac}^\perp\notag\\-2h^{ab}h^{cd}K(\partial_a,\partial_c)\biggl(g(R(N,\partial_b)N,\partial_d)+g(\nabla_{\partial_b}J^T,\partial_d)+g(\nabla_b N,\nabla_d N)\biggr)\label{thetwo}
\end{gather}
Where we have used eq. (\ref{theone}) and $L^{(J)}_{ab}$ is the first order Jacobi equation satisfied by $J=\nabla_NN$. This is consistent with any second order equation. One will get eq. (\ref{thetwo}) as the second order Jacobi equation when there is no perturbation. Next we will include the effect of perturbations and check how eq. (\ref{thetwo}) changes.

\subsubsection{Second order Jacobi equation with perturbation}
 When metric perturbations are included our definition of first order equation changes to $L_{ab}=B_{ab}-D_{ab}+S^{(N)}_{ab}+\accentset{(1)}{C}(\partial_a,\partial_b)+h^{cd}\accentset{(1)}{P}(K(\partial_a,\partial_b),\partial_c)\partial_d$. Naturally, the second order deviation equation (eq.\eqref{devi1}) will now include derivatives of perturbations. The new terms will come from derivative of $C(\partial_a,\partial_b)$ 
\begin{gather}
h^{ab}\nabla_N \accentset{(1)}{C}(\partial_a,\partial_b)^\perp=h^{ab}U_{ab}^\perp-h^{ab}g(\accentset{(1)}{C}(\partial_a,\partial_b)^\perp,(\nabla_{\partial_c}N)^\perp)h^{cd}\partial_d-(\nabla_E N)^\perp,
\end{gather}
where $E:=h^{ab}\accentset{(1)}{C}(\partial_a,\partial_b)^T$ and $U_{ab}:=\bigg[(\nabla_N \accentset{(1)}{C})(\partial_a,\partial_b)+ \accentset{(1)}{C}(\nabla_{\partial_a}N,\partial_b)+\accentset{(1)}{C}(\nabla_{\partial_a}N,\partial_b)\bigg]$
As before the derivative of the normal term is obtained by subtracting the derivative of the tangent part from the derivative of the total term. The covariant derivative of the final term gives,
\begin{gather}
h^{ab}\nabla_N\bigg(h^{cd}K(\partial_a,\partial_c)\accentset{(1)}{P}(\partial_b,\partial_d)\bigg)=h^{ab}h^{cd}\accentset{(1)}{P}(\partial_b,\partial_d)\bigg(L^{\perp}_{ac}-g\bigl(K(\partial_a,\partial_c),(\nabla_m N)^\perp\bigr)h^{mn}\partial_n\bigg)\notag\\
+h^{cd}h^{ab}K(\partial_a,\partial_c)\biggl(2g(\accentset{(1)}{C}(N,\partial_b),\partial_d)+\accentset{(1)}{P}(\nabla_{\partial_b}N,\partial_d)+\accentset{(1)}{P}(\partial_b,\nabla_{\partial_d}N)\biggr)
\end{gather}
Now we will provide the effect of action of $\delta_p$ on the terms. First we will check the action on $L_{ab}$. We will tabulate the action of $\delta_p$ on various terms contained in $L_{ab}$ such as $B_{ab},~D_{ab}$ etc.
\begin{gather}
h^{ab}\delta_P B_{ab}=\accentset{(1)}{C}(N,\nabla_{\partial_a}\partial_b)+\nabla_N \accentset{(1)}{C}(\partial_a,\partial_b)=:h^{ab}\mathcal M_{ab},~~
h^{ab}\delta_P B_{ab}^T
=h^{ab}\mathcal M_{ab}^T +h^{ab}P(B_{ab}^\perp,\partial_c)h^{cd}\partial_d
\end{gather}
The variation of the normal component therefore takes the form,
\begin{gather}
h^{ab}\mathcal M_{ab}^\perp -h^{ab}P(B_{ab}^\perp,\partial_c)h^{cd}\partial_d.
\end{gather}
Similarly one can obtain the variation of the other terms,
\begin{gather}
h^{ab}\delta_P D_{ab}=h^{ab}\mathcal Q_{ab}^\perp-h^{ab}P(D_{ab}^\perp,\partial_c)h^{cd}\partial_d,
\end{gather}
where $h^{ab}\mathcal Q_{ab}:=\nabla_E N+h^{ab}\accentset{(1)}{C}(N,(\nabla_{\partial_a}\partial_b)^T)$ and $E=P(\nabla_{\partial_a}\partial_b,\partial_c)h^{cd}\partial_c+g(\accentset{(1)}{C}(\partial_a,\partial_b),\partial_c)h^{cd}\partial_c-h^{ce}h^{df}g(\nabla_{\partial_a}\partial_b,\partial_c)P(\partial_e,\partial_f)\partial_c$. The variation of the $S_{ab}$ term is algebraically a bit more tedious to compute and yields,
\begin{gather}
h^{ab}\delta_P S_{ab}=-2h^{ab}h^{ce}h^{df}P(\partial_e,\partial_f)K(\partial_a,\partial_c)g(N,K(\partial_b,\partial_d))\notag\\
+2h^{ab}h^{cd}\bigg[\accentset{(1)}{C}(\partial_a,\partial_c)^\perp-h^{mn}P(K(\partial_a,\partial_c),\partial_m)\partial_n\bigg]g(N,K(\partial_b,\partial_d))\notag\\
-h^{ab}h^{cd}K(\partial_a,\partial_c)\bigg[2P(\nabla_bN,\partial_d)+2g(C(\partial_b,N),\partial_d)\bigg]
\end{gather}
The variation of the $\accentset{(1)}{C}(\partial_a,\partial_b)$ term gives
\begin{gather}
h^{ab}\delta_P \accentset{(1)}{C}(\partial_a,\partial_b)^\perp=\accentset{(2)}{C}(\partial_a,\partial_b)^\perp-2 P(\accentset{(1)}{C}(\partial_a,\partial_b))^\perp-h^{ab}P(\accentset{(1)}{C}(\partial_a,\partial_b)^\perp,\partial_c)h^{cd}\partial_d.
\end{gather}
Note that we have introduced a notation $\accentset{(2)}{C}(\partial_a,\partial_b)$ which is nothing but the expression for $C(\partial_a,\partial_b)$ with $\accentset{(1)}{P}(\partial_a,\partial_b)$ replaced by $\accentset{(2)}{P}(\partial_a,\partial_b)$. The final term yields,
\begin{gather}
h^{ab}\delta_P\bigg(h^{cd}K(\partial_a,\partial_c)P(\partial_b,\partial_d)\bigg)\notag\\
=h^{ab}\bigg[-h^{ce}h^{df}P(\partial_e,\partial_f)K(\partial_a,\partial_c)P(\partial_b,\partial_d)+h^{cd}\bigg(\accentset{(1)}{C}(\partial_a,\partial_c)^\perp-P(K(\partial_a,\partial_c),\partial_m)h^{mn}\partial_n\bigg)P(\partial_b,\partial_d)\notag\\
+h^{cd}K(\partial_a,\partial_c)\accentset{(2)}{P}(\partial_b,\partial_d)\bigg]
\end{gather}
The variation of the $h^{ab}$ is trivial and gives $\delta_P h^{ab}L_{ab}^\perp=-h^{ae}h^{bf}P(\partial_e,\partial_f)L_{ab}^\perp$. Througout this calculation we have used identitites given in appendix (\ref{Iden}). 

Further, we have used the fact that $[N,\partial_a]=0$. However if we didn't assume this then several additional tangent contribution would arise. However it will be clearfrom the next appendix  that these tangent terms won't contribute to the final equation.

\section{Tangent Variations}\label{tanvar}
In the following two sections we will derive certain identities for tangent variations of the extrinsic curvature and mean curvature vector $H$. These will be used in the main text in section \ref{sop}.
\subsection{Tangent variation of extrinsic curvature}\label{tv}
Let us start by calculationg the tangent variation of the extrinsic curvature of $\mathscr S$. Consider a vector $X \in \mathcal T_p\mathcal S$ such that $[X,\partial_a]\neq 0$. Note that, $\mathscr S$ being an integral sub manifold $[X,\partial_a]^\perp=0$. After a few algebraic steps it follows that,
\begin{gather}
\nabla_X K(\partial_a,\partial_b)=\bigg(R(X,\partial_a)\partial_b^\perp+(\nabla_{\partial_a}\nabla_{\partial_b}X)^\perp-(\nabla_{\nabla_{(\partial_a}\partial_b)^T}X)^\perp+g((\nabla_{\partial_a}\partial_b)^\perp,(\nabla_{\partial_c}X)^\perp)h^{cd}\partial_d \notag\\
+K(\partial_a,[\partial_b,X])+K([X,\partial_a],\partial_b)\bigg)
\end{gather}

\subsection{Tangent variation of $H$}\label{TVH}
From the above expression eq. (\ref{tv}), it follows that $h^{ab}\bigg[\nabla_{\partial_a}\nabla_{\partial_b}X+R(X,\partial_a)\partial_b-\nabla_{(\nabla_{\partial_a}\partial_b)^T}X \bigg]^\perp-2h^{ae}h^{bf}g(\nabla_e X,\partial_f)K(\partial_a,\partial_b)=\nabla_X H$. However there is another elegant way to show it using Gauss's equation.
\begin{gather}
h^{ab}\bigg[\nabla_{\partial_a}\nabla_{\partial_b}X+R(X,\partial_a)\partial_b-\nabla_{(\nabla_{\partial_a}\partial_b)^T}X \bigg]^\perp-2h^{ae}h^{bf}g(\nabla_e X,\partial_f)K(\partial_a,\partial_b)\notag\\
=h^{ab}\bigg[\nabla_{\partial_a}^\perp K(\partial_b,X)+K(\partial_a,D_{\partial_b}X)
+\nabla_{X}^\perp K(\partial_a,\partial_b)-K(\partial_a,\nabla_X\partial_b)-K(\nabla_X\partial_a,\partial_b)\notag\\
-\nabla_{\partial_a}^\perp K(X,\partial_b)+K(X,\nabla_{\partial_a}\partial_b)+K(\nabla_{\partial_a}X,\partial_b)-K((\nabla_{\partial_a}\partial_b)^T,X) \bigg]^\perp-2h^{ab}K(\partial_a,\nabla_{\partial_b} X)=\nabla_{X} H
\end{gather} 
\section{Further Identities}\label{Iden}

\begin{itemize}
\item The metric variation of the Riemann tensor can be calculated using the above identities and is given as,
\begin{gather}
\delta_P R(W,U)V=\accentset{(1)}{C}(W,\nabla_U V)+\nabla_W \accentset{(1)}{C}(U,V)-\accentset{(1)}{C}(U,\nabla_W V)-\nabla_U \accentset{(1)}{C}(W,V)-\accentset{(1)}{C}([W,U],V)
\end{gather}
\item The metric variation of the extrinsic curvature is given by,
\beq
h^{ab}\delta_p K(\partial_a,\partial_b)=\accentset{(1)}{C}(\partial_a,\partial_b)^\perp
\eeq
\item The following identity will be useful in the calculation of second order perturbation equation.
 \begin{gather}
h^{ab}[N,(\nabla_{\partial_a}\partial_b)^T]=h^{ab}[N,g((\nabla_{\partial_a}\partial_b),\partial_c)h^{cd}\partial_d]=h^{ab}\nabla_{N}\left[g((\nabla_{\partial_a}\partial_b),\partial_c)h^{cd}\right]\partial_d\notag\\
=g(B_{ab}-\nabla_{(\nabla_{\partial_a}\partial_b)^T}N,\partial_c)h^{cd}\partial_d
\end{gather}
\item The Codazzi Identity is given by
\begin{gather}
 Ric(N^T,N^{\perp})=h^{ab}g((\nabla_{N^T}K)(\partial_a,\partial_b),N^{\perp})-h^{ab}g(D_{\partial_a}W_{N^{\perp}}(N^T),\partial_b)\nonumber\\
 +h^{ab}g(K(\partial_{a},N^T),{\nabla^{\perp}}_{\partial_{b}}N^{\perp})+h^{ab}g(D_{\partial_{a}}N^T,W_{N^{\perp}}(\partial_{b}))
\end{gather}
\item The Gauss equation is given as
\begin{gather}
 -Ric(N^T,N^T)+h^{ab}g(K(\partial_{a},N^T),K(\partial_{b},N^T))=-\bar{Ric}(N^T,N^T)
\end{gather}
\item An identity for the change in the affine connection $\accentset{(1)}{C}$ is,
\begin{gather}
 2h^{cd}h^{ab}g(\accentset{(1)}{C}(\partial_a,\partial_c),\partial_d)g(N,\partial_b)=h^{cd}g(\accentset{(1)}{C}(\partial_c,N^T),\partial_d)
\end{gather}
\item To extract a surface term from a $\accentset {(1)}{P}$ dependent term one needs the following identity,
\begin{gather}
h^{cd}P(\partial_{c},\partial_{d})g(\nabla_{\partial_{a}}N,\partial_{b})=\nabla_{\partial_a}[h^{cd}P(\partial_c,\partial_d)g(N,\partial_b)]\nonumber\\-h^{cd}\Biggl[2P((\nabla_{\partial_a}\partial_c)^{\perp},\partial_d)+2g(\accentset{(1)}{C}(\partial_a,\partial_c),\partial_d)\Biggr]g(N,\partial_b)-h^{cd}P(\partial_{c},\partial_{d})g(N,\nabla_{\partial_{a}}\partial_{b})
\end{gather}
\end{itemize}


\begin{thebibliography}{99}
 \bibitem{nielsen2005geometric}
  Nielsen,~ Michael~ A,
  arXiv preprint quant-ph/050207

\bibitem{Jefferson:2017sdb} 
  R.~Jefferson and R.~C.~Myers,
  JHEP {\bf 1710}, 107 (2017)
  doi:10.1007/JHEP10(2017)107
  [arXiv:1707.08570 [hep-th]].



\bibitem{Susskind:2014rva} 
  L.~Susskind,
  [Fortsch.\ Phys.\  {\bf 64}, 24 (2016)]
  Addendum: Fortsch.\ Phys.\  {\bf 64}, 44 (2016)
  doi:10.1002/prop.201500093, 10.1002/prop.201500092
  [arXiv:1403.5695 [hep-th], arXiv:1402.5674 [hep-th]].



\bibitem{Brown:2015bva} 
  A.~R.~Brown, D.~A.~Roberts, L.~Susskind, B.~Swingle and Y.~Zhao,
  Phys.\ Rev.\ Lett.\  {\bf 116}, no. 19, 191301 (2016)
  doi:10.1103/PhysRevLett.116.191301
  [arXiv:1509.07876 [hep-th]].



\bibitem{Belin:2018fxe} 
  A.~Belin, A.~Lewkowycz and G.~Sárosi,
  Phys.\ Lett.\ B {\bf 789}, 71 (2019)
  doi:10.1016/j.physletb.2018.10.071
  [arXiv:1806.10144 [hep-th]].



\bibitem{Belin:2018bpg} 
  A.~Belin, A.~Lewkowycz and G.~Sárosi,
  JHEP {\bf 1903}, 044 (2019)
  doi:10.1007/JHEP03(2019)044
  [arXiv:1811.03097 [hep-th]].



\bibitem{Alishahiha:2015rta} 
  M.~Alishahiha,
  Phys.\ Rev.\ D {\bf 92}, no. 12, 126009 (2015)
  doi:10.1103/PhysRevD.92.126009
  [arXiv:1509.06614 [hep-th]].



\bibitem{bender2013advanced}
C.~M. Bender and S.~A. Orszag, \emph{Advanced mathematical methods for
  scientists and engineers I: Asymptotic methods and perturbation theory}.
  Springer Science \& Business Media, 2013.

\bibitem{lin1988mathematics}
C.-C. Lin and L.~A. Segel, \emph{Mathematics applied to deterministic problems
  in the natural sciences}, vol.~1. Siam, 1988.

\bibitem{Bhattacharya:2012mi} 
  J.~Bhattacharya, M.~Nozaki, T.~Takayanagi and T.~Ugajin,
  Phys.\ Rev.\ Lett.\  {\bf 110}, no. 9, 091602 (2013)
  doi:10.1103/PhysRevLett.110.091602
  [arXiv:1212.1164 [hep-th]].



\bibitem{Mishra:2015cpa} 
  R.~Mishra and H.~Singh,
  JHEP {\bf 1510}, 129 (2015)
  doi:10.1007/JHEP10(2015)129
  [arXiv:1507.03836 [hep-th]].



\bibitem{Quijada:2017zif} 
  E.~Quijada and H.~Boschi-Filho,
  arXiv:1711.08505 [hep-th].


\bibitem{Bhattacharya:2019zkb} 
A.~Bhattacharya, K.~T.~Grosvenor and S.~Roy,
arXiv:1905.02220 [hep-th].


\bibitem{Lewkowycz:2018sgn} 
  A.~Lewkowycz and O.~Parrikar,
  JHEP {\bf 1805}, 147 (2018)
  doi:10.1007/JHEP05(2018)147
  [arXiv:1802.10103 [hep-th]].



\bibitem{Mosk:2017vsz} 
  B.~Mosk,
  Class.\ Quant.\ Grav.\  {\bf 35}, no. 4, 045013 (2018)
  doi:10.1088/1361-6382/aaa4e9
  [arXiv:1710.01316 [hep-th]].



\bibitem{Bao:2019bib} 
  N.~Bao, C.~Cao, S.~Fischetti and C.~Keeler,
  arXiv:1904.04834 [hep-th].



\bibitem{Engelhardt:2019hmr} 
  N.~Engelhardt and S.~Fischetti,
  arXiv:1904.08423 [hep-th].



\bibitem{Speranza:2019hkr} 
  A.~J.~Speranza,
  arXiv:1904.08012 [gr-qc].



\bibitem{Ghosh:2017ygi} 
  A.~Ghosh and R.~Mishra,
  Phys.\ Rev.\ D {\bf 97}, no. 8, 086012 (2018)
  doi:10.1103/PhysRevD.97.086012
  [arXiv:1710.02088 [hep-th]].

\bibitem{Ryu:2006ef} 
  S.~Ryu and T.~Takayanagi,
  JHEP {\bf 0608}, 045 (2006)
  doi:10.1088/1126-6708/2006/08/045
  [hep-th/0605073].

\bibitem{Stanford:2014jda} 
  D.~Stanford and L.~Susskind,
  Phys.\ Rev.\ D {\bf 90}, no. 12, 126007 (2014)
  doi:10.1103/PhysRevD.90.126007
  [arXiv:1406.2678 [hep-th]].

\bibitem{Alishahiha:2015rta} 
  M.~Alishahiha,
  Phys.\ Rev.\ D {\bf 92}, no. 12, 126009 (2015)
  doi:10.1103/PhysRevD.92.126009
  [arXiv:1509.06614 [hep-th]].
  
\bibitem{Ghosh:2016fop} 
  A.~Ghosh and R.~Mishra,
  Phys.\ Rev.\ D {\bf 94}, no. 12, 126005 (2016)
  doi:10.1103/PhysRevD.94.126005
  [arXiv:1607.01178 [hep-th]].



\bibitem{Stewart:1974uz} 
  J.~M.~Stewart and M.~Walker,
  Proc.\ Roy.\ Soc.\ Lond.\ A {\bf 341}, 49 (1974).
  doi:10.1098/rspa.1974.0172



\bibitem{Hubeny:2007xt} 
  V.~E.~Hubeny, M.~Rangamani and T.~Takayanagi,
  JHEP {\bf 0707}, 062 (2007)
  doi:10.1088/1126-6708/2007/07/062
  [arXiv:0705.0016 [hep-th]].



\bibitem{Nozaki:2013vta} 
  M.~Nozaki, T.~Numasawa, A.~Prudenziati and T.~Takayanagi,
  Phys.\ Rev.\ D {\bf 88}, no. 2, 026012 (2013)
  doi:10.1103/PhysRevD.88.026012
  [arXiv:1304.7100 [hep-th]].



\bibitem{Bhattacharya:2013bna} 
  J.~Bhattacharya and T.~Takayanagi,
  JHEP {\bf 1310}, 219 (2013)
  doi:10.1007/JHEP10(2013)219
  [arXiv:1308.3792 [hep-th]].



\bibitem{Wall:2012uf} 
  A.~C.~Wall,
  Class.\ Quant.\ Grav.\  {\bf 31}, no. 22, 225007 (2014)
  doi:10.1088/0264-9381/31/22/225007
  [arXiv:1211.3494 [hep-th]].



\bibitem{Fonda:2014cca} 
  P.~Fonda, L.~Giomi, A.~Salvio and E.~Tonni,
  JHEP {\bf 1502}, 005 (2015)
  doi:10.1007/JHEP02(2015)005
  [arXiv:1411.3608 [hep-th]].



\bibitem{Ben-Ami:2016qex} 
  O.~Ben-Ami and D.~Carmi,
  JHEP {\bf 1611}, 129 (2016)
  doi:10.1007/JHEP11(2016)129
  [arXiv:1609.02514 [hep-th]].



\bibitem{Karar:2019bwy} 
  S.~Karar, R.~Mishra and S.~Gangopadhyay,
  arXiv:1904.13090 [hep-th].

\bibitem{Cavini:2019wyb} 
  G.~Cavini, D.~Seminara, J.~Sisti and E.~Tonni,
  arXiv:1907.10030 [hep-th].
  
\bibitem{Carmi:2015dla} 
  D.~Carmi,
  JHEP {\bf 1512}, 043 (2015)
  doi:10.1007/JHEP12(2015)043
  [arXiv:1506.07528 [hep-th]].
  
\bibitem{Faulkner:2015csl} 
  T.~Faulkner, R.~G.~Leigh and O.~Parrikar,
  JHEP {\bf 1604}, 088 (2016)
  doi:10.1007/JHEP04(2016)088
  [arXiv:1511.05179 [hep-th]].
   
   \bibitem{maple}
 Maple (13.0). Maplesoft, a division of Waterloo Maple Inc., Waterloo, Ontario.
  
  \bibitem{mathematica}
  Wolfram Research, Inc., Mathematica, Version 11.0.1.0, Champaign, IL (2016)
  
  
  \bibitem{grtensor}
  Peter~Musgrave, Denis~Pollney and Kayll~Lake, 
          GRTensorII software (Version 1.79 (R4), 2001). Queen's University, Kingston, Ontario, Canada.
  \end{thebibliography}
\end{document}